\documentclass[11pt]{article}
\usepackage{slashed}
\usepackage{amsfonts}
\usepackage{amsmath}
\usepackage{amssymb}
\usepackage{epsfig}

\usepackage{pgf,tikz}
\usetikzlibrary{arrows}
\usetikzlibrary{patterns}

\newtheorem{Theorem}{Theorem}

\newtheorem{Lemma}[Theorem]{Lemma}

\newcommand{\unit}{{\rm 1\hspace*{-0.4ex}%
\rule{0.1ex}{1.52ex}\hspace*{0.2ex}}}

\def\oz{\overline{z}}

\newcommand{\Z}{{\mathbb Z}}
\newcommand{\I}{{\mathbb I}}

\def\ov{\overline}
\def\ds{\displaystyle}
\def\sc{\scriptstyle}

\def\qed{\hfill$\blacksquare$}

\begin{document}

\title{\protect {\Large \textsc{Plethystic Vertex Operators and Boson-Fermion Correspondences}}}

\author{Bertfried Fauser
\footnote{Fachbereich Physik, Univers\"{a}t Konstanz, 78457 Konstanz, Germany \texttt{(Bertfried.Fauser@uni-konstanz.de)} } , 
Peter D Jarvis
\footnote{School of Physical Sciences (Mathematics and Physics), University of Tasmania, Tas 7001 Australia \texttt{(peter.jarvis@utas.edu.au)}} 
and Ronald C King
\footnote{School of Mathematics, University of Southampton, Southampton SO17 1BJ, UK \texttt{(r.c.king@soton.ac.uk)}    }}



\maketitle

\abstract
We study the algebraic properties of plethystic vertex operators, introduced 
in J. Phys. A: Math. Theor. {\bf 43} 405202 (2010), underlying the structure of symmetric functions associated with certain generalized universal character rings of subgroups of the general linear group, defined to stabilize tensors of Young symmetry type characterized by a partition of arbitrary shape $\pi$. 
Here we establish an extension of the well-known boson-fermion correspondence involving Schur functions
and their associated (Bernstein) vertex operators: for each $\pi$, the modes generated by the plethystic vertex operators and their suitably constructed duals, satisfy the anticommutation relations of a complex Clifford algebra.
The combinatorial manipulations underlying the results involve exchange identities exploiting the
Hopf-algebraic structure of certain symmetric function series and their plethysms.


\section{Introduction}\label{sec:Introduction}
The fundamental role of vertex operators and associated mathematical structures in the physics and geometry of string theory and two dimensional conformal field theory has long been recognised \cite{borcherds1986vertex, frenkel1988vertex}. In `free field' realizations the context is algebraic-combinatorial, {expressed} in terms of the structure and properties of the universal ring of symmetric functions, and it is this setting with which we are concerned in this note. Phrased in this language, 
{the origins of vertex operators} 
can be traced to the study of certain basic endomorphisms (see for example \cite{macdonald:1979a}). 
{Vertex} 
operators are also called Bernstein operators (in the terminology of Zelevinsky \cite{zelevinsky:1981b}). 

In recent work we have studied the relationship between symmetric functions and character rings for groups beyond the classical general linear, orthogonal and symplectic cases developed from the foundational works of Weyl \cite{weyl:1930a} (as expounded in the classic text of Littlewood \cite{littlewood:1940a}). Our systematic analysis \cite{fauser:jarvis:2003a, fauser:jarvis:king:wybourne:2005a, fauser:jarvis:king:2010d} (technicalities of which will be covered as needed in the following) allows progress to be made on what Littlewood referred to as \emph{restricted groups}: matrix subgroups of the general linear group which stabilise a fixed, but arbitrary, numerical tensor of given structure and Young symmetry type. For symmetry type specified by partition $\pi$, and for tensors of order $p=|\pi|$ ({the} number of tensor indices, or {rank}), 
we refer to these, possibly trivial, groups as $H_\pi$ subgroups of $GL(n)$ in dimension $n$. ({The} 
complex orthogonal and symplectic groups being the cases where {$|\pi|=2$ and} $\pi$ is symmetric 
and antisymmetric for even $n$, respectively, {each with an associated nonsingular bilinear form}).  

Our results from \cite{fauser:jarvis:2003a, fauser:jarvis:king:wybourne:2005a, fauser:jarvis:king:2010d, fauser:jarvis:king:2010b} are as follows. For each $\pi$ we have identified in the ring of symmetric functions, the linear basis of symmetric functions of type $\pi$ which plays the role of the universal character ring for the group $H_\pi$,
analogously to the way in which the standard basis of Schur functions, and Schur functions of symplectic and orthogonal type, provide the structure of the universal character rings in the general linear, orthogonal and symplectic cases (for the Hopf structure of the character rings in the latter cases see also \cite{fauser2012hopf}). A key technical aspect throughout \cite{fauser:jarvis:king:wybourne:2005a} is the exploitation of the symmetric function operation of \emph{plethysm}, as used by Littlewood \cite{littlewood:1940a}, and the systematic way in which (for example) plethysm distributes over symmetric function operations such as multiplication and skew. Specifically we have described the generalized $\pi$-Newell-Littlewood product rule, and the character branching rules for basis transformation between Schur functions and symmetric functions of $\pi$-type. In contrast to the case of the classical groups, a complete characterization of the Hopf algebraic structure in the form of primitive elements, units, and counits obviously depends not only on the partition $\pi$ but also on the specific $\pi$ tensor (for example, its tensor rank and perhaps a canonical form) and so is not possible to give {this complete characterization} in generic form. In particular, the universal characters will in general be indecomposable rather than irreducible, and so their utility requires further case-by-case consideration. On the other hand, it is quite feasible to explore general structural aspects of the new classes of symmetric functions, and this is the path which we follow here.

In \S~\ref{sec:PiSfns} below, we summarize the notation and background required to define the $\pi$-type symmetric functions, and in \S~\ref{sec:PiVertexDualPiVertex} as in~\cite{fauser:jarvis:king:2010d}, we establish the formal $\pi$-analogue of the standard (Bernstein) vertex operator modal product realization of the original Schur functions themselves. To complete the analogy, this leads to consideration of the appropriate `dual' vertex operators in the $\pi$ case. The main result of this paper is that the modes of the $\pi$-vertex operators, and those of their suitably constructed duals, together satisfy the anticommutation relations of an infinite dimensional complex Clifford algebra, that is, the standard `free fermion' algebra
--  as in the well known case of the Bernstein operators themselves. Our finding thereby generalizes earlier work of Baker~\cite{baker1996vertex} on vertex operators and duals, for orthogonal and symplectic characters (see also Jing and Nie~\cite{jing2015vertex}), which correspond to the {rank 2} 
symmetric and antisymmetric cases $\pi = (2)$, and $\pi = (1^2)$, respectively. However, our results have much more general validity --  the free fermion algebra is valid not just for the one part partitions $\pi =(p)$ and their duals $\pi = (1^p)$ which directly generalize the orthogonal and symplectic cases with weight $|\pi|=p=2$ (see below for definitions), but it obtains for arbitrary partition shapes $\pi$. Detailed proofs are provided in {\S~\ref{sec:PiVertexDualPiVertex}, augmented by the Appendices~\ref{sec:Appendix A}
and~\ref{sec:Appendix B}}, via rearrangement lemmas for computing operator ordering, with use of the underlying symmetric function (co)algebraic structure.

{Just as $\pi$-Schur functions, $\smash{s^{(\pi)}_\lambda(X)}$,
 may be defined in terms of products of vertex operators acting on the identity, 
so we may {also} define dual $\pi$-Schur functions, $\smash{s^{*(\pi)}_\lambda(X)}$, in terms of products of our dual vertex operators acting on the identity. This is done in
\S~\ref{sec:PiSfnDualPiSfn}, in which the $\pi$-Schur functions and their duals are then evaluated in terms of ordinary Schur functions by exploiting 
a powerful normal ordering lemma proved in Appendix~\ref{sec:Appendix C}.}

The paper concludes in \S~\ref{sec:Conclusions} with a brief summary, a short survey of related work, and some discussion of implications of the results and future work.

\section{Symmetric functions of $\pi$-type}
\label{sec:PiSfns}
As is well known, in $1+1$-dimensional quantum field theory, field mode expansions in advanced and retarded variables $x\pm ct$, become in the Euclidean picture, Laurent expansions in a complex variable $z$. In the simplest (``chiral scalar'') case, the mode operators $\alpha_n$, $\alpha_{-n} \equiv \alpha_n{}^\dagger$ ($n\in {\mathbb Z}_{>0}$) fulfil the quantum mechanical commutation relations of an infinite Heisenberg algebra,
${[}\alpha_m,\alpha_n {]} = m\,\delta_{m\!+\!n,0}\,$, $m, n \in {\mathbb Z}_{\ne 0}\,$.
In the combinatorial equivalent, these operators are in turn realized on the ring $\Lambda(X)$ of symmetric functions of a countably infinite alphabet of indeterminates $X = \{ x_1,x_2, \cdots\}$\,, and it is in this setting that we wish to investigate the calculus of vertex operators. 

In order to develop this, we begin with some notational preliminaries (following \cite{macdonald:1979a}). The ring $\Lambda(X)$ has various distinguished algebraic generating sets, of which we will need the power sum symmetric functions $p_n(X)$, 
\[
p_n(X) =\sum_{k} x_k{}^n\,, 
\]
together with the complete and elementary symmetric functions, $h_n(X)$ and $e_n(X)$, defined via generating series as follows: 
\begin{eqnarray} 
M(z;X)\ &=\ \ds \prod_k\frac{1}{(1-z \,x_k )}\ =\ \sum_{n \ge 0}\, z^n\, h_n(X)   \,, \label{eqn:MzX}\\ 
L(z;X) = M(z;X)^{-1} &=\ \ds {\prod_k\ (1-z \,x_k ) }\ = \ \sum_{n \ge 0}\, (-z)^n\, e_n(X)\,, \label{eqn:LzX} 
\end{eqnarray}
where
\begin{equation}\label{eqn:hnen}
h_n(X) =  \hskip-1ex \sum_{k_1 \le k_2 \le \cdots \le k_n}\hskip-3ex x_{k_1} x_{k_2}\cdots x_{k_n}\, 
\quad\mbox{and}\quad    
e_n(X) =  \hskip-1ex \sum_{k_1 <  k_2 < \cdots < k_n} \hskip-3ex x_{k_1} x_{k_2}\cdots x_{k_n}\,.  
\end{equation}
{
In the special case $z=1$ we set $M(X)=M(1;X)$ and $L(X)=L(1;X)$, while for the sake of typographical 
simplicity, where $X$ is to be understood, we often write $M(z)$ and $L(z)$ for $M(z;X)$ and $L(z;X)$,
respectively.
}

Discussion of linear bases of $\Lambda(X)$ entails elements labelled by (integer) partitions $\lambda$. 
If $\lambda$ is a partition of $n$ we write $\lambda\vdash n$, and 
$\lambda=(\lambda_1,\lambda_2,\ldots)$ is a sequence of 
non-negative integers $\lambda_i$ for $i=1,2,\ldots $\,,
such that $\lambda_1\geq\lambda_2\geq\cdots\geq0$, 
with $\lambda_1+\lambda_2+\cdots=n$. The partition   
$\lambda$ is said to be of weight $|\lambda|=n$ and length
$\ell(\lambda)$ where $\lambda_i>0$ for all $i\leq\ell(\lambda)$ 
and $\lambda_i=0$ for all $i>\ell(\lambda)$. In specifying $\lambda$
the trailing zeros, that is those parts $\lambda_i=0$, are often 
omitted, while repeated parts are sometimes written in exponent form
$\lambda=(\cdots,2^{m_2},1^{m_1})$ where $\lambda$ contains $m_i$
parts equal to $i$ for $i=1,2,\ldots$. For each such partition,
$n(\lambda)=\sum_{i=1}^n (i-1)\lambda_i$ and 
$z_\lambda=\prod_{i\geq1} i^{m_i}\, m_i!$.

{
A convenient graphical visualization of a partition $\lambda$ 
is via~\cite{macdonald:1979a} a Young diagram $F^\lambda$, consisting of $|\lambda|$ boxes arranged in $\ell(\lambda)$ left adjusted
rows of lengths $\lambda_1,\lambda_2,\ldots,\lambda_{\ell(\lambda)}$ from top to bottom. Given two partitions $\kappa$ and $\lambda$, we write $\kappa\subseteq\lambda$ if and only if $F^\lambda$ contains $F^\kappa$, that is to say $\kappa_i\le \lambda_i$ for all $i\le \ell(\kappa)$, a circumstance of great importance for defining the operation of skew (see below). In the sequel, in the context of dual symmetic functions, we shall also require the partition $\lambda'$ conjugate to $\lambda$, with parts $\lambda_1',\lambda_2',\ldots$ that specify the number of boxes in the columns of $F^\lambda$ from left to right (whose Young diagram $F^{\lambda'}$ is the reflection of
$F^\lambda$ in the diagonal). It follows that $|\lambda'|=|\lambda|$, and $\ell(\lambda')=\lambda_1$.
}

Returning to the definitions, it is the case that the complete and elementary symmetric functions are particular members of the
linear basis of Schur functions $s_\lambda(X)$ corresponding to the specific circumstance of one row or one column 
{Young diagrams, respectively, that is} 
$h_m(X) = s_{(m)}(X)$, and $e_m(X) = s_{(1^m)}(X)$\,. Their 
expansions {(\ref{eqn:hnen})} are 
special instances of the following combinatorial definition of the Schur functions $s_\lambda(X)$ for partitions $\lambda$ of arbitrary shape.
Let ${\cal T}^\lambda$ denote the set of semistandard tableaux $T$ of shape $\lambda$ with entries from $\{1,2,\ldots,n\}$, and 
let $X^T=x_1^{\#1}x_2^{\#2}\cdots x_n^{\#n}$ where $\#k$ is the number of entries $k$ in $T$. Then we have simply
\begin{align}
\label{eq:SfnDefn}
s_\lambda(X) = &\,\sum_{T \in  {\mathcal T}^\lambda} X^T\,.
  \end{align}

The ring $\Lambda(X)$ is given the status of a Hilbert space by defining an inner product such that the Schur functions are an orthonormal basis.
The associative product in the ring (the outer product, $f\cdot g(X) = f(X)g(X)$) then has a natural adjoint, the operation of symmetric function skew, uniquely defined by duality as 
\[
\langle\, f/g\,\, |\, h\rangle = \langle f \,|\, g\!\cdot\! h\,\rangle\,,
\]
where the skew product (or quotient) is variously written as $f/g =D(g)f = g^\perp f$ with the latter form adopted for typographical convenience
in much of what follows.
{Given two Schur functions $s_\lambda$ and $s_\kappa$, the skew Schur function $s_{\lambda/\kappa}:=s_\lambda/s_\kappa = 
s_\kappa^\perp s_\lambda$ is 
non-zero if and only if $\kappa\subseteq\lambda$.}
The adjoint arises in the realization of the afore-mentioned infinite Heisenberg algebra, in that negative integer-indexed modes $\alpha_{-k}$ are associated with power sums $p_k$ (operating by point multiplication), with their adjoints $D(p_k)$ identified with the $\alpha_k$. Acting on symmetric functions expressed in terms of the algebraic basis of power sum functions, one has indeed (up to scaling) the standard Schr\"{o}dinger representation $D(p_k) = k \partial/\partial p_k$\,.

An important further operation on symmetric functions needed in the sequel is that of plethysm, defined as follows. Suppose $f(X)$ has an expansion in monomials in $X$, $f(X) = \sum_i y_i$. Adopt these monomials as elements of a new {countably} infinite alphabet $Y=\{y_1,y_2,\cdots\}$\,. Then for any Schur function $s_\lambda(X) $, the plethysm of $f$ by $s_\lambda$, $s_\lambda{[}f{]}(X):= s_\lambda(Y)$, is the symmetric function of the composite alphabet (also denoted  $f \otimes s_\lambda$)\,. We will need the plethysms of the basic complete and elementary symmetric functions by the series $M(X)$, which in the weight 2 case are 
\begin{eqnarray} 
M_{(2)}(z;X) = M(z; {s_{(2)}(X)}) = \prod_{i\le j}\frac{1}{(1-zx_ix_j)}= {\sum_{r \ge 0}\, z^r\, s_{(r)}[s_{(2)}](X)  } =\sum_{r \ge 0} z^r s_{(2)} \otimes h_r(X) \,; \nonumber \\
M_{(1^2)}(z;X)) = M(z;{s_{(1^2)}(X)}) = \prod_{i< j}\frac{1}{(1-zx_ix_j)}= {\sum_{r \ge 0}\, z^r\,s_{(r)}[s_{(1^2)}](X)    } =\sum_{r \ge 0} z^r s_{(1^2)}\otimes h_r(X)\,. \nonumber
\end{eqnarray}
with inverses $L_{(2)}(z;X)=L(z;{s_{(2)}(X)})$ {and} $L_{(1^2)}(z;X))=L(z;{s_{(1^2)}(X)})$\,.

The connection with character theory, and the generalizations leading to the vertex operators which we wish to introduce, are as follows. For a finite alphabet, the celebrated Schur functions are well known to give the characters of finite dimensional irreducible representations of $GL(n)$, where the parameters $(x_1,x_2,\cdots, x_n)$ are the eigenvalues of an $n\times n$ 
(invertible) complex matrix. In the inductive limit, the ring $\Lambda(X)$ and indeed the Schur functions as linear basis embody the corresponding universal character ring. Remarkably, in the case of the classical orthogonal and symplectic groups, the universal character ring is still carried by $\Lambda(X)$, with the basis of irreducible characters, the Schur functions of orthogonal type and symplectic type, respectively defined by skewing by series (setting $z {=}1$),
\begin{equation}
s^O_\lambda(X) = {s_\lambda^{(2)}(X)}= s_\lambda/L_{(2)}(X)   \quad\mbox{and}\quad   
s^{Sp}_\lambda(X) ={s_\lambda^{(1^2)}(X)} = s_\lambda/L_{(1^2)}(X)\,. \nonumber 
\end{equation}
While this invertible mapping amounts to a linear automorphism of the space $\Lambda(X)$, it is not isometric, and key structural elements of $\Lambda(X)$ as a Hopf algebra in the case of the orthogonal and symplectic groups, differ from those for $\Lambda(X)$ for the general linear group universal character ring~\cite{fauser2012hopf}.

In recent work~\cite{fauser:jarvis:king:wybourne:2005a}
we have investigated the natural extension of the above to the case of formal universal character rings of putative matrix subgroups $H_\pi$ of the general linear group which leave invariant a fixed numerical tensor of Young symmetry type associated with a partition $\pi$ of arbitrary tensor weight, with the classical orthogonal and symplectic groups being the above $|\pi|=2$ cases $(2)$ and $(1^2)$, respectively. For example, if $\pi=(3)$ the corresponding symmetric functions of type $\pi=(3)$ will be defined by $s^{(3)}_\lambda(X) = s_\lambda/L_{(3)}(X)$, and a given Schur function will be expressed as $s_\lambda(X) = s^{(3)}_\lambda/M_{(3)}(X)$, where
\begin{eqnarray} 
M_{(3)}( {X}) = \prod_{i\le j\le k}\frac{1}{(1- x_ix_j x_k)}\,,\nonumber 
\end{eqnarray}
as a sum of symmetric functions of this type corresponding to the branching rule from a module of the general linear group, to (generically indecomposable) modules of the $H_{(3)}$ subgroup. In view of the definition of plethysm, and (\ref{eq:SfnDefn}) above,
we have in general
\begin{align}
    M_\pi(z;X)&= \prod_{T\in{\cal T}^\pi} \frac{1}{1-z\,X^T} = \sum_{r\geq0} z^r\, s_{(r)}[s_\pi](X) \label{eqn:Mpi}\,; \\ \cr
    L_\pi(z;X)&= \prod_{T\in{\cal T}^\pi} (1-z\,X^T) = \sum_{r\geq0} (-1)^r\, z^r\, s_{(1^r)}[s_\pi](X) \label{eqn:Lpi}\,, 					
\end{align}
and, for arbitrary $\pi$, we have 
\begin{align}
s^{(\pi)}_\lambda(X) &= s_\lambda/L_{\pi}(X)= L_{\pi}^\perp(X)s_\lambda\,,  \qquad
s_\lambda(X) = s^{(\pi)}_\lambda/M_{\pi}(X)= M_{\pi}^\perp(X)s^{(\pi)}_\lambda\,,  \label{eqn:SpiLambdaDef}
\end{align}
being the definition, and generalized branching rule, respectively. 

We refer to our paper~\cite{fauser:jarvis:king:wybourne:2005a} for further details of the $\pi$-type symmetric functions, which as we have emphasized, provide a natural
extension of the symmetric function theory of Littlewood~\cite{littlewood:1940a}, as developed for the character rings of the classical groups, 
to formal character rings associated to an arbitrary partition $\pi$. For example, in addition to the above branching rule, there exists~\cite{fauser:jarvis:king:wybourne:2005a} a generalization of the Newell-Littlewood~\cite{newell:1951a, littlewood:1958b} formula extending the outer product of orthogonal and symplectic characters, to the outer product of {$H_\pi$} 
characters (again corrresponding to generically indecomposable modules). Several special cases, involving such product and branching rules, with associated dimension formulae, are given in~\cite{fauser:jarvis:king:wybourne:2005a} by way of illustration.

\section{$\pi$-vertex operators, dual $\pi$-vertex operators and exchange relations}
\label{sec:PiVertexDualPiVertex}
Central to the role of Schur functions in various contexts is the ability to compute them in different ways. As mentioned in the introduction, they can be identified as matrix elements of basic endomorphisms, products of the so-called Bernstein or vertex operators. Recall the mutually inverse series $M(z;X)$ and $L(z;X)$. Using Taylor's series for the natural logarithm, and suppressing the underlying alphabet, we have
\begin{eqnarray}
M(z) = \exp \left( \sum_{n=1}^\infty\frac{z^n}{n} p_n\right)\,,
\qquad  L(z^{-1}) = \exp \left( - \sum_{n=1}^\infty \frac{z^{-n}}{n} p_{n}\right)\,, \nonumber
\end{eqnarray}
in terms of which the vertex operator is defined
\begin{eqnarray}
\label{eqn:VOdef}
V(z) = M(z) D\big(L(z^{-1})\big) \equiv \exp \left( \sum_{n=1}^\infty\frac{z^n}{n} p_n\right)
\exp \left( - \sum_{n=1}^\infty  {z^{-n}}\frac{\partial}{\partial p_{n}}\right)\,.
\end{eqnarray}
The Schur functions themselves can be recovered in terms of modal projections of products of strings of these objects acting on the identity, namely
\begin{eqnarray}
\label{eqn:SfnVO}
s_\lambda = {[}Z^\lambda{]}\, V(z_1)\cdot V(z_2)\cdots V(z_k)\cdot 1
\end{eqnarray}
where the notation $[z^m] \cdots $ selects the coefficient of the appropriate power in a series expansion, with ${[}Z^\lambda{]}:= [z_1^{\lambda_1}][z_2^{\lambda_2}]\cdots[z_k^{\lambda_k} ] $\, the corresponding multinomial coefficient.
In the case of Schur functions, the same $s_\lambda$ can be obained equivalently via
strings of both the basic vertex operators above, as well as the so-called `dual' vertex operators
\begin{eqnarray}
\label{eqn:DualVOdef}
V^*(w) = L(w) D\big(M(w^{-1})\big) \equiv \exp \left(-\sum_{n=1}^\infty\frac{w^n}{n} p_n\right)
\exp \left(+ \sum_{n=1}^\infty  {w^{-n}}\frac{\partial}{\partial p_{n}}\right)\,,
\end{eqnarray}
acting on the identity.
We shall not need further combinatorial details, but integral to the construction is the remarkable fact that the complex modes of the vertex and dual vertex operators fulfil the elementary algebraic relations of an algebra once again intimately related to quantum field expansions, but for the ``chiral fermion'' case rather than the chiral scalar case corresponding to the modes of the Heisenberg algebra. The results are most simply stated in terms of an expanded algebra which includes the additional zero mode operator $\alpha_0$, augmented by its canonical conjugate {$q$, such that} $[q,\alpha_0] = i\unit$\,. {We define} 
the full vertex operators, including contributions from the zero modes, as
\begin{equation}
\label{eqn:fullXVO}
X(z) = V(z) e^{iq} z^{\alpha_0}
\quad\mbox{and}\quad  
X^*(z) = V^*(z) z^{-\alpha_0}e^{-iq}\,.
\end{equation}

The crucial property is the following. The full vertex operators $X(z)$, $X^*(z)$ fulfil the operator product exchange relations:
\begin{align}
X(z)X(w) + X(w)X(z)=&\,0\,;\nonumber \\
X^*(z)X^*(w) + X^*(w)X^*(z) =&\,0\,;\nonumber \\
X(z)X^*(w) + X^*(w)X(z)=&\,\delta_{z,w}\,\unit\,,
\label{eqn:StandardXrelations}
\end{align}
{where} 
$\delta_{z,w}$ (with Laurent expansion $\sum_{n=-\infty}^{+\infty} (z/w)^n$) is the distributional $\delta$-function, and $\unit$ is the unit operator. Correspondingly, the modes $X_m$ and $X^*_n$, $m,n\in \Z$  under Laurent expansions of $X(z)$ and $X^*(z)$, defined by
\begin{equation}
\label{eqn:Laurent}
X(z) = \sum_{n\in {\mathbb Z}} z^{n+{\alpha_0}} X_n \quad\mbox{and}\quad X^*(z) = \sum_{n\in {\mathbb Z}} z^{-n-{\alpha_0}} X^*_{n}\,,
\end{equation}
satisfy  the free fermion algebraic anticommutation relations of an infinite-dimensional complex Clifford algebra,
\begin{equation}
\{X_m,X_n\} = \,0\,, \qquad \{X^*_m,X^*_n\} = \,0\,,\qquad \{X_m,X^*_n\} = \,\delta_{m+n,0}\,\unit \,, \quad\mbox{for all $m,n \in {\mathbb Z}$}\,.
\label{eqn:StandardXmoderelations}
\end{equation}

Our interest here is in extending this structure to symmetric functions of type $\pi$\,, along the lines already explored by Baker \cite{baker1996vertex} for vertex operators for the aforementioned Schur functions of orthogonal and symplectic type (see also Jing \cite{jing2015vertex}). In \cite{fauser:jarvis:king:2010d} we gave the definition of $\pi$-vertex operators, {$V_\pi(z)$}, constructed to have the same property with respect to the $\pi$-Schur functions, {$\smash{s^{(\pi)}_\lambda}$}, as the standard vertex operators in (\ref{eqn:VOdef}) 
have to standard Schur functions in (\ref{eqn:SfnVO})\,, namely
\begin{eqnarray}
s^{(\pi)}_\lambda = {[}Z^\lambda{]}\ V_\pi(z_1) V_\pi(z_2)\cdots V_\pi(z_k)\cdot 1\,,\qquad\mbox{where} \label{eqn:sfnpiV}   \\
V_\pi(z) = (1-\delta_{\pi,(p)}z^p)\, M(z)\, L^\perp({z}^{-1})\,\left.\prod\right._{k=1}^{p-1}L^\perp_{\pi/(k)}(z^{k})\,.
\label{eqn:Vpidef}
\end{eqnarray}
{In order to extend our analysis to obtain}
the complete set of exchange relations between the $\pi$-vertex operators
{it is necessary to introduce}
suitably constructed dual vertex operators $V^{*}_\pi(z)$, 
{and then to adjoin} to both {$V_{\pi}(z)$ and $V^{*}_\pi(z)$} 
zero mode contributions as {as is done} in the standard case, to give the corresponding full vertex operators $X^\pi(z)$ {and} $X^{*\pi}(z)$.

With these preliminaries we now turn to the statement of the main result of this paper, in which for simplicity we write
$\ov{z}=z^{-1}$:
 
\begin{Theorem}\label{thm:main}
{
For each partition $\pi$ and any $z$ let
\begin{eqnarray}
  V_\pi(z)&:=& \ds M(z)\, L^\perp(\ov{z})\, \prod_{k>0} \, L^\perp_{\pi/(k)}(z^k)\,; \label{eqn:Vpi}\\ 
	V^*_\pi(z)&:=& \ds L(z)\, M^\perp(\ov{z})\, \prod_{k\ge 0}\, M^\perp_{\pi/(1^{2k+1})}(z^{2k+1}) \, \prod_{k>0}\, L^\perp_{\pi/(1^{2k})}(z^{2k})\,, \label{eqn:V*pi}
\end{eqnarray}
where it is to be understood that all the Schur functions in $M(w), L(w), M^\perp(w)$ and $L^\perp(w)$, for any $w$, depend on the same sequence of indeterminates
$(x_1,x_2,\ldots)$ whose specification, again for the sake of simplicity, has been suppressed.
}

Furthermore, let the associated full vertex operators $X^\pi(z)$ and $X^{*\pi}(z)$, constructed by adjoining zero mode contributions in the usual way, 
be defined as in (\ref{eqn:Laurent}) above by
\begin{equation}
X^\pi(z) = V_\pi(z) e^{iq} z^{\alpha_0}:= \sum_{n\in {\mathbb Z}} z^{n+{\alpha_0}} X^\pi_{-n} \quad\mbox{and}\quad 
X^{*\pi}(z) = V_\pi^*(z) z^{-\alpha_0}e^{-iq}:= \sum_{n\in {\mathbb Z}} z^{-n-{\alpha_0}} X^{*\pi}_{{n}}\,. 
\end{equation}
\noindent
{Then} we have
\begin{description}
\item[(a)] For all $\pi$, 
$X^\pi(z)$ {and} $X^{*\pi}(z)$ satisfy 
\begin{align}
X^\pi(z) X^\pi(w) + X^\pi(w) X^\pi(z)=&\,0\,;\nonumber \\
X^\pi(z) X^\pi(w) + X^\pi(w) X^\pi(z)=&\,0\,;\nonumber \\
X^\pi(z) X^\pi(w) + X^\pi(w) X^\pi(z)=&\,\unit \,\delta_{z,w}\,. 
\label{eqn:GenericXpiX*piRelations}
\end{align}
\item[(b)] The modes $X^{\pi}_m$ {and} $X^{*\pi}_n$ 
fulfil the free fermion anticommutation relations of a complex Clifford algebra:
\begin{align}\label{eqn:XpimXpinClifford}
\{ X^{\pi}_m, X^{\pi}_n\} = \,0 \,; \ \ 
 \{ X^{*\pi}_m , X^{*\pi}_n\}= \,0 \,; \ \ 
\{ X^{\pi}_m, X^{*\pi}_n\} =\, \delta_{m+n,0} \,{\unit}\ \quad\mbox{for all $m,n \in {\mathbb Z}$}\,, 
\end{align}
where $\{\cdot\,,\,\cdot\}$ signifies an anticommutator.

\end{description}
\end{Theorem}

{It should be noted in the definitions (\ref{eqn:Vpi}) and (\ref{eqn:V*pi})
that although the summations over $k$ are in principle unbounded, the expressions are finite by virtue of the fact that
\begin{equation}\label{eqn:MLequal0}
    M_{\pi/\kappa}(z) = M^\perp_{\pi/\kappa}(z) = 1 \quad\mbox{and}\quad L_{\pi/\kappa}(z)=L^\perp_{\pi/\kappa}(z)=1 \quad\mbox{for all $\kappa\not\subseteq\pi$\,.} 	
\end{equation}
since $s_{\pi/\kappa}=0$ if $\kappa\not\subseteq\pi$.
Moreover, to recover (\ref{eqn:Vpidef}) from (\ref{eqn:Vpi}) it is only necessary to note in addition 
the second of the following two identities that will play a crucial role in what follows:
\begin{equation}\label{eqn:MLequal1}
    M_{(0)}(z)=M^\perp_{(0)}(z) = \frac{1}{1-z} \quad\mbox{and}\quad L_{(0)}(z)=L^\perp_{(0)}(z)= (1-z)\,. 
\end{equation}
Each of these is an immediate consequence of the fact that $s_{(0)}=1$ for any sequence of suppressed parameters
$(x_1,x_2,\ldots)$.
}

{Below we provide} a proof of these results, drawing heavily on a basic lemma proven in {Appendix~\ref{sec:Appendix A}.}
This gives the relations for 4 types of reshufflings of the $M$ and $L$ plethystic factors and their adjoints, which make up the individual  exponential components contributing to the full vertex operators, all of which are required for ordering vertex operator products. 

By way of illustration of the reordering formalism, note that the vertex operators $V$, $V^*$ themselves are defined as `normal ordered' products $\cong e^P e^D$, where $P$ is the series linear in power sums $p_k$ and in the differential realization, $D$ is a polynomial in derivatives $\partial/\partial p_k$ of degree $p-1$ for $|\pi|=p\ge 2$\,. Algebraic relations are in turn derived by reducing products to normal ordered form, and comparing terms (c.f. \cite{fauser:jarvis:king:2010d}). The following well known exponential adjoint identity is applicable,
\begin{eqnarray}
\label{eqn:CauchyExpFormula}
e^D e^P = e^P \big(e^{-P} e^D e^P\big) = e^P e^{\big(D + {[}D,P{]} + {\frac 12} {[}{[}D,{]}P{]},P{]} 
+\cdots \big)}\,,
\end{eqnarray}
and in fact requires only $p$ commutator terms, up to ${[} {[}\!\cdot\!\cdot\!\cdot\!{[}D, P{]},\cdot\!\cdot\!\cdot{]},P{]}/(p-1)!$\,, {which, in view of the structure of $D$, must be a numerical scalar with no dependence on the $p_k$.} However, in view of the complexity of the evaluations required using this method (which would entail intermediate computations with Schur functions in the power sum basis), the {reordering} lemma instead exploits the underlying Hopf structure and its compatibility with canonical operations such as plethysm and skew (see {Appendix}~\ref{sec:Appendix A} for details). \\

\noindent
\textbf{Proof of Theorem~\ref{thm:main}}:\\

Noting that $L^\perp(\ov{z})=L^\perp_{(1)}(\ov{z})$ and $L^\perp_{(0)}(\ov{z}w)=(1-\ov{z}w)$, it follows from the repeated use of (\ref{eqn:LpiperpM}) that
\begin{align}
  &V_\pi(z)\, V_\pi(w) =  M(z)\, L^\perp_{(1)}(\ov{z})\, \prod_{i>0}\, L^\perp_{\pi/(i)}(z^i) \ 
	                            M(w)\, L^\perp(\ov{w})\, \prod_{j>0}\, L^\perp_{\pi/(j)}(w^j) \cr
							&= M(z)\, M(w)\, L^\perp_{(1)}(\ov{z}) \, L^\perp_{(0)}(\ov{z}w)\,   
							\prod_{i>0}\prod_{k\ge 0} L^\perp_{\pi/((i)(k))}(z^iw^k)\ 
							     L^\perp(\ov{w})\, \prod_{j>0}\, L^\perp_{\pi/(j)}(w^j) \cr
							&=	(1-\ov{z}w)\	P_{\pi}(z,w) \,,	\label{eqn:VpiVpi}												
\end{align}
where 
\begin{equation}
       P_{\pi}(z,w)= M(z)\ M(w)\ L^\perp(\ov{z})\ L^\perp(\ov{w}) \ \prod_{i,j\ge 0:(i,j)\neq(0,0)} L^\perp_{\pi/((i)(j))}(z^iw^j) \,, \label{eqn:P}
\end{equation}
from which it can be seen that $P_{\pi}(w,z)=P_{\pi}(w,z)$.

Similarly, with the use of (\ref{eqn:MpiperpL}) and (\ref{eqn:LpiperpL}) 
\begin{align}
 			&V^\ast_\pi(z) V^\ast_\pi(w) = L(z)\, M^\perp_{(1)}(\ov{z})\, \prod_{i\ge 0}\, M^\perp_{\pi/(1^{2i+1})}(z^{2i+1})\
			                        \prod_{i>0}\, L^\perp_{\pi/(1^{2i})}(z^{2i}) \cr
										&~~~~~~~~~~~~~~~~~~~ L(w)\, M^\perp(\ov{w})\, \prod_{j\ge 0}\, M^\perp_{\pi/(1^{2j+1})}(w^{2j+1})\ 
			                        \prod_{j> 0}\, L^\perp_{\pi/(1^{2j})}(w^{2j}) \cr			
				&=	L(z)\	L(w)\ M^\perp_{(1)}(\ov{z})\ L^\perp_{(0)}(\ov{z}w) \, M^\perp(\ov{w})\ \prod_{j\ge 0} M^\perp_{\pi/(1^{2j+1})}(w^{2j+1})\ 
			                        \prod_{j>0} L^\perp_{\pi/(1^{2j})}(w^{2j}) \cr	
				&~~~~ \prod_{i\ge 0}\prod_{k\ge 0} M^\perp_{\pi/((1^{2i+1})(1^{2k}))}(z^{2i+1}w^{2k})\ 			
				  \prod_{i\ge 0}\prod_{k\ge 0} L^\perp_{\pi/((1^{2i+1})(1^{2k+1}))}(z^{2i+1}w^{2k+1})\ \cr  
				&~~~~ \prod_{i>0}\prod_{k\ge 0} L^\perp_{\pi/((1^{2i})(1^{2k}))}(z^{2i}w^{2k})\ 			
				  \prod_{i>0}\prod_{k\ge 0} M^\perp_{\pi/((1^{2i})(1^{2k+1}))}(z^{2i}w^{2k+1})\ \cr  
        &=	(1-\ov{z}w)\	Q_{\pi}(z,w)\,, \label{eqn:V*piV*pi}																																		
\end{align}
where
\begin{align}
       Q_{\pi}(z,w) &= L(z)\ L(w)\ M^\perp(\ov{z})\ M^\perp(\ov{w}) \cr
						& \prod_{i,j\ge 0} M^\perp_{\pi/((1^{2i+1})(1^{2j}))}(z^{2i+1}w^{2j})\ 			
				      \prod_{i,j\ge 0} L^\perp_{\pi/((1^{2i+1})(1^{2j+1}))}(z^{2i+1}w^{2j+1})\ \cr  
				    & \prod_{i,j\ge 0:(i,j)\neq(0,0)} L^\perp_{\pi/((1^{2i})(1^{2j}))}(z^{2i}w^{2j})\ 			
				      \prod_{i,j\ge 0} M^\perp_{\pi/((1^{2i})(1^{2j+1}))}(z^{2i}w^{2j+1})\,,  									
						\label{eqn:Q} 
\end{align}
with $Q_{\pi}(z,w)=Q_{\pi}(w,z)$.

In the case $w\neq z$, the repeated use of (\ref{eqn:LpiperpL}) yields
\begin{align}
 			&V_\pi(z)\, V^\ast_\pi(w) =  M(z)\, L^\perp_{(1)}(\ov{z})\, \prod_{i>0}\, L^\perp_{\pi/(i)}(z^i) \cr
										&~~~~~~~~~~~~~~~~~~~~~ L(w)\, M^\perp(\ov{w})\, \prod_{j\ge 0}\, M^\perp_{\pi/(1^{2j+1})}(w^{2j+1})\ 
			                        \prod_{j>0}\, L^\perp_{\pi/(1^{2j})}(w^{2j}) \cr			
				&=	M(z)\,	L(w)\, L^\perp_{(1)}(\ov{z})\, M^\perp_{(0)}(\ov{z}w)\, M^\perp(\ov{w})\, \prod_{j\ge 0}\, M^\perp_{\pi/(1^{2j+1})}(w^{2j+1})\ 
			                        \prod_{j>0}\, L^\perp_{\pi/(1^{2j})}(w^{2j})  \cr	
				&~~~~ \prod_{i>0}\prod_{k\ge 0}\, L^\perp_{\pi/((i)(1^{2k}))}(z^{i}w^{2k})\ 			
				  \prod_{i>0}\prod_{k\ge 0}\, M^\perp_{\pi/((i)(1^{2k+1}))}(z^{i}w^{2k+1})\ \cr  
        &=	(1-\ov{z}w)^{-1}\	R_{\pi}(z,w)\,, \label{eqn:VpiV*pi}																																		
\end{align}
where
\begin{align}
       R_{\pi}(z,w)&= M(z)\, L(w)\, L^\perp_{(1)}(\ov{z})\, M^\perp_{(1)}(\ov{w}) \cr 
					 	&~~~~ \prod_{i,j\ge 0:(i,j)\neq(0,0)} L^\perp_{\pi/((i)(1^{2j}))}(z^{i}w^{2j})\ 			
				      \prod_{i,j\ge 0} M^\perp_{\pi/((i)(1^{2j+1}))}(z^{i}w^{2j+1})\,.  
				    	\label{eqn:R} 
\end{align}
Finally, by using (\ref{eqn:MpiperpM}) and (\ref{eqn:LpiperpM}) one obtains
\begin{align}
      V^\ast_\pi(z) V_\pi(w) &= L(z)\, M^\perp_{(1)}(\ov{z})\, \prod_{i\ge 0} M^\perp_{\pi/(1^{2i+1})}(z^{2i+1})\ 
			                        \prod_{i>0}\, L^\perp_{\pi/(1^{2i})}(z^{2i}) \cr			
			  &~~~~~~~~~~	M(w)\, L^\perp(\ov{w})\, \prod_{j>0} L^\perp_{\pi/(j)}(w^j) \cr											
				&=	L(z)\,	M(w)\, M^\perp_{(1)}(\ov{z})\, M^\perp_{(0)}(\ov{z}w)\, 	L^\perp(\ov{w})\, \prod_{j>0}\, L^\perp_{\pi/(j)}(w^j) \cr 
				&~~~~~~~~~~ \prod_{i\ge 0}\prod_{k\ge 0} M^\perp_{\pi/((1^{2i+1})(k))}(z^{2i+1}w^{k})\ 			
				            \prod_{i>0}\prod_{k\ge 0}\, L^\perp_{\pi/((1^{2i})(k))}(z^{2i}w^{k})\cr
				& =	(1-\ov{z}w)^{-1}\	S_{\pi}(z,w)\,, \label{eqn:V*piVpi}																																		
\end{align}
where
\begin{align}
       S_{\pi}(z,w) &= L(z)\, M(w)\, M^\perp(\ov{z})\, L^\perp(\ov{w}) \cr 
					  &~~~~ \prod_{i,j\ge 0}\, M^\perp_{\pi/((1^{2i+1})(j))}(z^{2i+1}w^{j})\ 			
				      \prod_{i,j\ge 0:(i,j)\neq(0,0)}\, L^\perp_{\pi/((1^{2i})(j))}(z^{2i}w^{j})\,. 									
					 					         \label{eqn:S} 
\end{align}
Comparing (\ref{eqn:R}) and (\ref{eqn:S}) it can be seen that $S_{\pi}(w,z)=R_{\pi}(z,w)$, and 
in the special case $z=w$ we have
\begin{align}
       R_{\pi}(z,z) &= M(z)\, L(z)\, L^\perp(\ov{z})\, M^\perp(\ov{z})  \cr 
						 &~~~~ \prod_{i,j\ge 0}\, M^\perp_{\pi/((1^{2i+1})(j))}(z^{2i+j+1})\ \prod_{i,j\ge 0:(i,j)\neq(0,0)}\, L^\perp_{\pi/((1^{2i})(j))}(z^{2i+j})\,.  									
					 					         \label{eqn:Rpipi} 
\end{align}
However, it should be noted that
\begin{align}
    &\prod_{i,j\geq0}\, z^{2i+j+1}\, s_{(1^{2i+1})}(X) \, s_{(j)}(X) = \prod_{a,b\geq0}\, z^{a+b+1}\,s_{(a+1,1^b)}(X) \cr
\mbox{and}	\cr
		&\prod_{i,j\geq0:(i,j)\neq(0,0)}\, z^{2i+j}\, s_{(1^{2i})}(X) \, s_{(j)}(X) = \prod_{a,b\geq0}\, z^{a+b+1}\, s_{(a+1,1^b)}(X)
\end{align}
for all $X=(x_1,x_2,\ldots)$.
It follows that
\begin{align}
       &R_{\pi}(z,z) = M(z)\, L(z)\, L^\perp(\ov{z})\, M^\perp(\ov{z})\ 
						 \prod_{a,b\geq0}\, M^\perp_{\pi/(a+1,1^b)}(z^{a+b+1})\ L^\perp_{\pi/(a+1,1^b)}(z^{a+b+1})	\ = \ 1				
					 					         \label{eqn:Rpipi-hooks} 
\end{align}
where the last equality follows from the fact that $M_\sigma(w)$ and $L_\sigma(w)$ are mutually inverse series
for all $\sigma$ and all $w$. 

Turning to the full vertex operators, it follows from the above that
\begin{align}
    &X^\pi(z)\, X^\pi(w) + X^\pi(w)\, X^\pi(z) = V_\pi(z)\, V_\pi(w)\, e^{iq} z^{\alpha_0} e^{iq} w^{\alpha_0} + V_\pi(w)\, V_\pi(z)\, e^{iq} w^{\alpha_0} e^{iq} z^{\alpha_0} \cr
		&= (1-\ov{z}w)\, P_\pi(z,w)\, \ov{z}\,\ov{w}^2 (zw)^{\alpha_0} e^{2iq} + (1-\ov{w}z)\, P_\pi(w,z)\, \ov{w}\,\ov{z}^2 (wz)^{\alpha_0} e^{2iq}\cr
		&= ((z-w) + (w-z))\, P_\pi(z,w)\, \ov{zw}^2 (zw)^{\alpha_0} e^{2iq} \ =\ 0\,, 
\end{align}
where use has been made of (\ref{eqn:XXzw}) and the fact that $P_\pi(z,w)=P_\pi(w,z)$.
Similarly,
\begin{align}
    &X^{*\pi}(z)\, X^{*\pi}(w) + X^{*\pi}(w)\, X^{*\pi}(z) = V^*_\pi(z)\, V^*_\pi(w)\,z^{-\alpha_0} e^{-iq} w^{-\alpha_0} e^{-iq}  
		                    + V^*_\pi(w)\, V^*_\pi(z)\, w^{-\alpha_0} e^{-iq} z^{-\alpha_0}  e^{-iq}\cr
		&= (1-\ov{z}w)\, Q_\pi(z,w)\, \ov{w} (zw)^{-\alpha_0} e^{-2iq} + (1-\ov{w}z)\, Q_\pi(w,z)\, \ov{z} (wz)^{-\alpha_0} e^{-2iq}\cr
		&= ((z-w) + (w-z))\, Q_\pi(z,w)\, (zw)^{-\alpha_0} e^{-2iq} \ =\ 0\,, 
\end{align}
where use has been made of (\ref{eqn:X*X*zw}) and the fact that $Q_\pi(z,w)=Q_\pi(w,z)$.
Furthermore, we have
\begin{align}
    &X^{\pi}(z)\, X^{*\pi}(w) + X^{*\pi}(w)\, X^{\pi}(z) = V_\pi(z)\, V^*_\pi(w)\, e^{iq} z^{\alpha_0} w^{-\alpha_0} e^{-iq}  
		                    + V^*_\pi(w)\, V_\pi(z)\, w^{-\alpha_0} e^{-iq} e^{iq} z^{\alpha_0} \cr
		&= (1-\ov{z}w)^{-1}\, R_\pi(z,w)\, \ov{z}w (z\ov{w})^{\alpha_0} + (1-\ov{w}z)^{-1}\, S_\pi(w,z)\, (z\ov{w})^{\alpha_0} \cr
		&= \ds \left\{\,\frac{\ov{z}w}{1-\ov{z}w} + \frac{1}{1-\ov{w}z}\,\right\} R_\pi(z,w)\, (z\ov{w})^{\alpha_0}\,, 
\end{align}
where use has been made of (\ref{eqn:XX*zw}) and (\ref{eqn:X*Xwz}), together with the fact that $S_\pi(w,z)=R_\pi(z,w)$.
For $z\neq w$ the factor in braces is well defined and equals $0$, but diverges to $+\infty$ in the limit $z/w\rightarrow  1$. This can be seen by 
noting that in view of the origin of the terms $1/(1-\ov{z}w)$ and $1/(1-\ov{w}z)$, namely $M_{(0)}(\ov{z}w)$ and $M_{(0)}(\ov{w}z)$, respectively, 
we can write this factor in the form
\begin{equation}
   \frac{\ov{z}w}{1-\ov{z}w} + \frac{1}{1-\ov{w}z} = \sum_{k\ge 1} (w/z)^k + \sum_{k\ge 0} (z/w)^k = \sum_{k\in \Z} (z/w)^k = \delta_{z,w}
\end{equation}
where the identification with a $\delta$ function has been argued elsewhere (see for example~\cite{frenkel2001vertex}). 
It then follows that
\begin{align}
        X^{\pi}(z)\, X^{*\pi}(w) + X^{*\pi}(w)\, X^{\pi}(z) 
					&= \delta_{z,w}\ R_\pi(z,w)\, (z\ov{w})^{\alpha_0} = \delta_{z,w}\ R_\pi(z,z) = \delta_{z,w} \unit\,, 
\end{align}
as required to complete the proof of (\ref{eqn:GenericXpiX*piRelations}).
The validity of (\ref{eqn:XpimXpinClifford}) then follows from the simultaneous Laurent expansions of (\ref{eqn:GenericXpiX*piRelations})
with respect to $z$ and $w$.
\qed
\bigskip

\noindent{\bf Example}:
To complete this section we offer a direct demonstration of the free fermion result for the test case $\pi = (3)$, 
{for which it follows from the definitions (\ref{eqn:Vpi}) and (\ref{eqn:V*pi}) that}
\begin{equation}
   V_{(3)}(z) = M(z) L^\perp(\ov{z}) L^\perp_{(2)}(z) L^\perp_{(1)}(z^2) L^\perp_{(0)}(z^3)
	\quad\mbox{and}\quad
	 V^*_{(3)}(z) = L(z) M^\perp(\ov{z}) L^\perp_{(2)}(z)  \,. \nonumber
\end{equation}
Beginning with the product $V_{(3)}(z)V_{(3)}(w)$, we have
\begin{align}
V_{(3)}(z) \, V_{(3)}(w)  &=  
         M(z) L^\perp(\ov{z}) L^\perp_{(2)}(z) L^\perp_{(1)}(z^2) L^\perp_{(0)}(z^3) \cdot
         M(w) L^\perp(\ov{w}) L^\perp_{(2)}(w) L^\perp_{(1)}(w^2) L^\perp_{(0)}(w^3) \cr 
				     &= M(z) M(w) \cdot L^\perp(\ov{z}) L^\perp_{(0)}(\ov{z}w) \cdot L^\perp_{(2)}(z)L^\perp_{(1)}(zw)L^\perp_{(0)}(zw^2) \cr
						 &~~~~~~ \cdot L^\perp_{(1)}(z^2)L^\perp_{(0)}(z^2w) \cdot L^\perp_{(0)}(z^3) \cdot L^\perp(\ov{w}) L^\perp_{(2)}(w) L^\perp_{(1)}(w^2) L^\perp_{(0)}(w^3) \cr
&= (1- \ov{z}w)(1-z^3)(1-w^3)(1-zw^2)(1-z^2w) \cr 
&~~~~~~ M(z) M(w) L^\perp(\ov{z}) L^\perp(\ov{w}) 
               L^\perp_{(2)}(z) L^\perp_{(2)}(w) L^\perp_{(1)}(zw) L^\perp_{(1)}(z^2) L^\perp_{(1)}(w^2)\,, \nonumber
\end{align}
where three applications of the $L^\perp M$ {reordering} rule (\ref{eqn:LpiperpM}) have been used to establish the normal-ordered form of the result.
Up to the 
factor $(1- \overline{z}w)$, the terms are symmetrical with respect to the interchange $z\leftrightarrow w$. 
Recalling the additional zero-mode {reordering} terms {(\ref{eqn:XXzw})} involved in the full vertex operators , we have 
\[
\{ X^{(3)}(z), X^{(3)}(w)\} =  0\,.
\]

Turning to $V^*_{(3)}(z)V^*_{(3)}(w)$, we note similarly {that using two applications of the $M^\perp L$ reordering 
rule (\ref{eqn:MpiperpL}) gives}
\begin{align}
V^*_{(3)}(z)\,V^*_{(3)}(w)&= L(z) M^\perp(\ov {z}) M_{(2)}^\perp(z) \cdot  L(w) M^\perp(\ov {w})  M_{(2)}^\perp(w)  \cr
       &= L(z) L(w) \cdot M^\perp(\ov {z}) L^\perp_{(0)}(zw) \cdot M_{(2)}^\perp(z) L_{(1)}^\perp(zw) 
			            \cdot M^\perp(\ov {w})  M_{(2)}^\perp(w) \cr
  &= (1- \ov {z}w) L(z) L(w) M^\perp{}(\ov {z}) M^\perp{}(\ov {w}) 
	             \cdot M_{(2)}^\perp(z) M_{(2)}^\perp{}(w) L_{(1)}^\perp(zw) ,.\nonumber 
\end{align}
Once again up to the 
factor $(1- \overline{z}w)$, the terms are symmetrical with respect to the interchange $z\leftrightarrow w$. 
Recalling the additional zero-mode {reordering} terms {(\ref{eqn:X*X*zw})} involved in the full vertex operators, 
we {can conclude that}
\[
\{ X^{*(3)}(z), X^{*(3)}(w)\} =  0\,.
\]

The reordering of the mixed products ${V}_{(3)}(z){V}^*_{(3)}(w)$ {and ${V}^*_{(3)}(w){V}_{(3)}(z)$ entails 
three applications of the $L^\perp L$ reordering rule (\ref{eqn:LpiperpL}) and two of the $M^\perp M$ reordering rule (\ref{eqn:MpiperpM}) 
as follows:}
\begin{align}
&V_{(3)}(z) V^*_{(3)}(w) =  M(z) L^\perp(\ov {z}) L^\perp_{(2)}(z) L^\perp_{(1)}(z^2) L^\perp_{(0)}(z^3) \cdot
                                     L(w) M^\perp{}(\ov {w})  M_{(2)}^\perp{}(w) \cr
                        &=   M(z) L(w) \cdot L^\perp(\ov {z}) M^\perp_{(0)}(\ov{z}w) 
												             \cdot L^\perp_{(2)}(z) M^\perp_{(1)}(zw) 
																		 \cdot L^\perp_{(1)}(z^2) M^\perp_{(0)}(z^2w) 
												\cdot  L^\perp_{(0)}(z^3) \cdot M^\perp{}(\ov {w})  M_{(2)}^\perp{}(w) \cr
												&= (1-z^3)(1-\ov{z}w)^{-1}(1-z^2w)^{-1} 
												\ M(z) L(w) L^\perp(\ov{z}) M^\perp{}(\ov {w})  L^\perp_{(2)}(z) L^\perp_{(1)}(z^2)
                                M^\perp_{(2)}(w) M^\perp_{(1)}(zw)\,; \nonumber \\[.1cm] 
&V^*_{(3)}(w) V_{(3)}(z) =  L(w) M^\perp(\ov {w})  M^\perp_{(2)}(w) \cdot M(z) L^\perp(\ov {z}) L^\perp_{(2)}(z) L^\perp_{(1)}(z^2)  L^\perp_{(0)}(z^3) \cr
                        &=  L(w) M(z) \cdot M^\perp(\ov{w})  M^\perp_{(0)}(\ov{w}z) 
												        \cdot M^\perp_{(2)}(w)  M^\perp_{(1)}(wz)  M^\perp_{(0)}(wz^2)
																\cdot L^\perp(\ov{z}) L^\perp_{(2)}(z) L^\perp_{(1)}(z^2)  L^\perp_{(0)}(z^3) \cr
												&=	(1-\ov{w}z)^{-1}(1-wz^2)^{-1}	(1-z^3) 
												\ L(w) M(z) M^\perp(\ov{w}) L^\perp(\ov {z})
												           M^\perp_{(2)}(w)  M^\perp_{(1)}(wz) L^\perp_{(2)}(z) L^\perp_{(1)}(z^2) 	\,. \nonumber
\end{align}
Inserting the normal ordering factors (\ref{eqn:XX*zw}) and (\ref{eqn:X*Xwz}) arising from the zero mode contributions to the full vertex operators, gives
\begin{align}
\{ X_{(3)}(z), X^*_{(3)}(w)\} &= 
\frac{1-z^3}{1-z^2w}\left\{ \frac{\ov {z}w}{1- \ov {z}w} + \frac{1}{1- \ov {w}z} \right\} \cr
 &~~~~~~~~~~~~~~~~\cdot M(z) L(w) L^\perp(\ov {z})  M^\perp{}(\ov {w})  
                        L^\perp_{(2)}(z) M^\perp_{(2)}(w) L^\perp_{(1)}(z^2) M^\perp_{(1)}(zw) \cdot (z/w)^{\alpha_0}\,. \nonumber
 \end{align}
Here the factor in braces $\{ \cdots \}$ can be recognised as the complex $\delta$ function $\delta_{w,z}$. In the limit $w\rightarrow z$, 
the first and final factors each reduce to 1, and the accompanying series and their adjoints all cancel in pairs
to give the unit operator. Thus finally we establish
\[
\{ X_{(3)}(z), X^*_{(3)}(w)\} = \delta_{w,z}\,\unit \,.
\]
whose modal equivalent is the claimed infinite complex Clifford algebra (\ref{eqn:XpimXpinClifford}).


\section{$\pi$-Schur functions and their duals}
\label{sec:PiSfnDualPiSfn}
In~\cite{fauser:jarvis:king:2010d} we established the expression (\ref{eqn:sfnpiV}) for $\pi$-Schur functions $\smash{s^{(\pi)}_\lambda}$
in terms of our vertex operators $V_\pi(z)$. In this section, having defined dual vertex operators $V^*_\pi(z)$, we explore how, in turn, their modal products may be used 
to define what can be called dual $\pi$-Schur functions $\smash{s^{*(\pi)}_\lambda}$. In both cases, we may exploit our reordering Lemma~\ref{lem:reordering} to identify alternative, more conventional expressions for both $\smash{s^{(\pi)}_\lambda}$ and $\smash{s^{*(\pi)}_\lambda}$ (in the case of $\smash{s^{(\pi)}_\lambda}$, recovering our known constructions~\cite{fauser:jarvis:king:2010d}).  To be precise, we have:

\begin{Theorem}\label{thm:spi-s*pi}
Let $\lambda$ be a partition of length $\ell(\lambda)=m$ and let $Z=(z_1,z_2,\ldots,z_m)$. Then for
any $X=(x_1,x_2,\ldots)$ and partition $\pi$, let
\begin{align}
   s^{(\pi)}_\lambda(X) &= [Z^\lambda]\ V_\pi(z_1;X) V_\pi(z_2;X) \cdots V_\pi(z_m;X)\cdot 1\,; \label{eqn:spiX} \\ \cr
	s^{*(\pi)}_\lambda(X) &= [Z^\lambda]\ V^*_\pi(z_1;X) V^*_\pi(z_2;X) \cdots V^*_\pi(z_m;X)\cdot 1\,. \label{eqn:s*piX} 
\end{align}
Then
\begin{align}
   s^{(\pi)}_\lambda(X) &= [s_\lambda(Z)]\ M(XZ)\ L_\pi(Z) \,; \label{eqn:spiML}   \\ \cr
	 s^{*(\pi)}_\lambda(X) &= \begin{cases} [s_\lambda(Z)]\ L(XZ)\ L_{\pi'}(Z)&\mbox{if $|\pi|$ is even};\cr
	                                [s_\lambda(Z)]\ L(XZ)\ M_{\pi'}(Z))&\mbox{if $|\pi|$ is odd}.\cr
											\end{cases}			\label{eqn:s*piLM}			
\end{align}
where $\pi'$ is the conjugate of $\pi$. {Moreover, we have
\begin{equation}\label{eqn:spi-Lperps}
     s^{(\pi)}_\lambda(X) = L^\perp_\pi(X)\, s_\lambda(X)
				\quad\mbox{and}\quad
		 s^{*(\pi)}_\lambda(X) = (-1)^{|\lambda|}\, L^\perp_{\pi}(X)\, s_{\lambda'}(X)\,,
\end{equation}
together with the inverse forms,
\begin{equation}\label{eqn:s-Mperpspi}
s_\lambda(X) = M^\perp_\pi(X) s^{(\pi)}_\lambda(X) \quad\mbox{and}\quad
{s_{\lambda'}(X) = (-1)^{|\lambda|} {M}^\perp_\pi(X) s^{*(\pi)}_{\lambda}(X)}\,.
\end{equation}}
In particular, we have the identification $s^{*(\pi)}_\lambda(X) = (-1)^{|\lambda|}s^{(\pi)}_{\lambda'}(X)$.

\end{Theorem} 

\noindent{\bf Proof}:
The first of these has been proved previously~\cite{fauser:jarvis:king:2010d}, but for completeness and as a guide to proving the second  
we offer a second self-contained derivation, based this time on the use of
Lemma~\ref{lem:VVV*V*}. This provides explicit formulae for the normal ordering of 
products of vertex operators and of products of their duals that generalise the products of pairs given in (\ref{eqn:VpiVpi})
and (\ref{eqn:V*piV*pi}). It follows directly from the first part, (\ref{eqn:VVV}), of this lemma that
\begin{align}
    &V_\pi(z_1) V_\pi(z_2) \cdots V_\pi(z_m)\cdot 1 \cr
		&~~~~~~~~ = \prod_{1\le i<j\le m} (1-\ov{z_i}z_j)\ \prod_{\ell=1}^m\, M(z_\ell)\, 
		                  \prod_{\begin{array}{c} \sc i_1,i_2,\ldots,i_m\geq 0\cr \sc (i_1,i_2,\ldots,i_m)\neq(0,0,\ldots,0) \cr \end{array}} 
													 (1-z_1^{i_1}z_2^{i_2}\cdots z_m^{i_m})^{ c^\pi_{(i_1)(i_2)\cdots(i_m)} } \cr
		&~~~~~~~~ = ~~~ Z^{-\delta}\ \prod_{1\le i<j\le m} (z_i-z_j)\ \prod_{\ell=1}^m\, M(z_\ell)\ L_\pi(Z)
													\end{align}
where $\delta=(m-1,\ldots,1,0)$ and $c^\pi_{(i_1)(i_2)\cdots(i_m)}$ is the generalised Littlewood-Richardson coefficient defined by
\begin{equation}
 s_{(i_1)}\, s_{(i_2)} \, \cdots \, s_{(i_m)} = \sum_{\pi}\, c^\pi_{(i_1)(i_2)\cdots(i_m)}\ s_\pi \,,
\end{equation}  
and (elaborating on (\ref{eqn:MLequal0}), (\ref{eqn:MLequal1})) use has been made of the fact that 
\[
L^\perp_\sigma(w)\cdot 1 = \begin{cases} 1-w&\mbox{if $\sigma=(0)$};\cr  1&\mbox{otherwise},\cr\end{cases}
\]
while
\[
   L_\pi(Z)= L_\pi(1;Z) = \prod_{T\in {\cal T}^\pi} (1-Z^T) = \prod_{i_1,i_2,\ldots,i_m\geq 0}\ (1 - z_1^{i_1}z_2^{i_2}\cdots z_m^{i_m})^{c^\pi_{(i_1)(i_2)\cdots(i_m)}} \,.
\]

It follows that 
\begin{align}
    s^{(\pi)}_\lambda(X) &= [Z^\lambda]\ V_\pi(z_1;X) V_\pi(z_2;X) \cdots V_\pi(z_m;X)\cdot 1 \cr
		          &= [Z^{\lambda+\delta}]\ \prod_{1\le i<j\le m} (z_i-z_j)\ \prod_{\ell=1}^m\, M(z_\ell;X)\ L_\pi(Z) \cr
							&= [s_\lambda(Z)] \ M(X,Z)\,L_\pi(Z) \,
\end{align}
where we have first restored the explicit dependence on some arbitrary $X=(x_1,x_2,\ldots)$,
recognised that 
\[
   \prod_{\ell=1}^m\, M(z_\ell;X) =\prod_{\ell=1}^m \prod_{k\geq1} (1-z_\ell x_k)^{-1} = M(XZ),
\] 
and made use of the fact that $s_\lambda(Z) = (Z^{\lambda+\delta} + \cdots ) / \prod_{1\le i<j\le m} (z_i-z_j)$ 
where all the terms represented by $\cdots$ are distinct from $Z^{\lambda+\delta}$.

Similarly, it follows from the second part, (\ref{eqn:V*V*V*}), of Lemma~\ref{lem:VVV*V*} that if $|\pi|$ is even then
in exactly the same way as above
\begin{align}
		 V^*_\pi(z_1) V^*_\pi(z_2) \cdots V^*_\pi(z_m) \cdot 1 &= \prod_{1\le i<j\le m} (1-\ov{z_i}z_j)\ \prod_{\ell=1}^m\, L(z_\ell)\,  \cr
		                 &~~~~ \cdot   \prod_{\begin{array}{c} \sc i_1,i_2,\ldots,i_m\geq 0\cr \end{array}} 
												 (1-z_1^{i_1}z_2^{i_2}\cdots z_m^{i_m})^{c^\pi_{(1^{i_1})(1^{i_2})\cdots(1^{i_m})}} \cr		
		             &~~~~~~~~ = ~~~ Z^{-\delta}\ \prod_{1\le i<j\le m} (z_i-z_j)\ \prod_{\ell=1}^m\, L(z_\ell)\ L_{\pi'}(Z) \,,
\end{align} 
 where the last step is a consequence of the conjugacy identity 
\[
    c^\pi_{(1^{i_1})(1^{i_2})\cdots(1^{i_m})} = c^{\pi'}_{(i_1)(i_2)\cdots(i_m)}\,.
\] 
Hence for $|\pi|$ even  
\begin{align}
    s^{(\pi)}_\lambda(X) &= [Z^\lambda]\ V^*_\pi(z_1;X) V^*_\pi(z_2;X) \cdots V^*_\pi(z_m;X)\cdot 1 \cr
		          &= [Z^{\lambda+\delta}]\ \prod_{1\le i<j\le m} (z_i-z_j)\ \prod_{\ell=1}^m\, L(z_\ell;X)\ L_{\pi'}(Z) \cr
							&= [s_\lambda(Z)] \ L(XZ)\,L_{\pi'}(Z) \,.
\end{align}

On the other hand if $|\pi|$ is odd then
\begin{align}
		 V^*_\pi(z_1) V^*_\pi(z_2) \cdots V^*_\pi(z_m) \cdot 1 &= \prod_{1\le i<j\le m} (1-\ov{z_i}z_j)\ \prod_{\ell=1}^m\, L(z_\ell)\,  \cr
		                 &~~~~ \cdot   \prod_{\begin{array}{c} \sc i_1,i_2,\ldots,i_m\geq 0\cr \end{array}} 
													 c^\pi_{(1^{i_1})(1^{i_2})\cdots(1^{i_m})} (1-z_1^{i_1}z_2^{i_2}\cdots z_m^{i_m})^{-1} \cr		
		             &~~~~~~~~ = ~~~ Z^{-\delta}\ \prod_{1\le i<j\le m} (z_i-z_j)\ \prod_{\ell=1}^m\, L(z_\ell)\ M_{\pi'}(Z) \,,
\end{align} 
since (again c.f. (\ref{eqn:MLequal0}), (\ref{eqn:MLequal1}))
\[
M^\perp_\sigma(w)\cdot 1 = \begin{cases} (1-w)^{-1}&\mbox{if $\sigma=(0)$};\cr  1&\mbox{otherwise},\cr\end{cases}
\]
while
\[
   M_{\pi'}(Z)= M_{\pi'}(1;Z) = \prod_{T\in {\cal T}^{\pi'}} (1-Z^T)^{-1} = \prod_{i_1,i_2,\ldots,i_m\geq 0}\ c^{\pi'}_{(i_1)(i_2)\cdots(i_m)}\ (1 - z_1^{i_1}z_2^{i_2}\cdots z_m^{i_m})\,.
\]
It follows that for $|\pi|$ odd  
\begin{align}
    s^{(\pi)}_\lambda(X) &= [Z^\lambda]\ V^*_\pi(z_1;X) V^*_\pi(z_2;X) \cdots V^*_\pi(z_m;X)\cdot 1 \cr
		          &= [Z^{\lambda+\delta}]\ \prod_{1\le i<j\le m} (z_i-z_j)\ \prod_{\ell=1}^m\, L(z_\ell;X)\ M_{\pi'}(Z) \cr
							&= [s_\lambda(Z)] \ L(XZ)\,M_{\pi'}(Z) \,.
\end{align}
This completes the proof of (\ref{eqn:spiML}) and (\ref{eqn:s*piLM}). 

To deal with the second pair of equalities, (\ref{eqn:spi-Lperps}), consider first the coefficients $\ell_{\pi\nu}$ and $m_{\pi\nu}$ defined 
by the expansions:
\begin{equation}
 L_\pi(Z)= \sum_{k\ge 0}\, (-1)^k s_{(1^k)}[s_\pi(Z)] = \sum_\nu\, \ell_{\pi\nu}\, s_\nu(Z)
\quad\mbox{and}\quad 
 M_\pi(Z)= \sum_{k\ge 0}\, s_{(k)}[s_\pi(Z)] = \sum_\nu\, m_{\pi\nu}\, s_\nu(Z)\,.
\end{equation}
Then from (\ref{eqn:spiML}) we have
\begin{align}
    s^{(\pi)}_\lambda(X) &=  [s_\lambda(Z)] \ M(XZ)\,L_\pi(Z) = [s_\lambda(Z)]\ \prod_\mu\, s_\mu(X)\, s_{\mu}(Z)\  \prod_\nu\, \ell_{\pi\nu}\, s_\nu(Z) \cr
		                     &=  [s_\lambda(Z)] \prod_{\mu,\nu}\, \ell_{\pi\nu}\,s_\mu(X)\, \prod_\kappa\, c_{\mu\nu}^\kappa\, s_\kappa(Z) 
												            = \prod_{\mu,\nu}\, \ell_{\pi\nu}\, c_{\mu\nu}^\lambda\, s_\mu(X) \cr
												 &=  \prod_{\nu}\, \ell_{\pi\nu}\,s_{\lambda/\nu}(X) =   L^\perp_\pi(X) s_\lambda(X)\,.
\end{align}
This completes the first part of (\ref{eqn:spi-Lperps}).

The second part is naturally more complicated and requires the use of 
Littlewood's theorem of conjugates of plethysms~\cite{littlewood:1944} which states that for all $\mu$, $\nu$ and $Z$ that
\begin{equation}
     (s_\mu[s_\nu](Z))'\  = \ \begin{cases}\ s_\mu[s_{\nu'}](Z)&\mbox{if $|\nu|$ is even};\cr
		                                   \ s_{\mu'}[s_{\nu'}](Z)&\mbox{if $|\nu|$ is odd}.\cr \end{cases}
\end{equation}
It follows that for $|\pi|$ even
\begin{align}
     L_{\pi'}(Z) &= \sum_{k\ge 0}\, (-1)^k s_{(1^k)}[s_{\pi'}](Z) = \sum_\nu \ell_{\pi'\nu'}\, s_{\nu'}(Z)\cr
		             &= \sum_{k\ge 0}\, (-1)^k (\,s_{(1^k)}[s_{\pi}](Z)\,)' = \sum_\nu \ell_{\pi\nu}\, s_{\nu'}(Z) \nonumber
\end{align}
so that $\ell_{\pi'\nu'}=\ell_{\pi\nu}$, while for $|\pi|$ odd
\begin{align}
     M_{\pi'}(Z) &= \sum_{k\ge 0}\, s_{(k)}[s_{\pi'}](Z) = \sum_\nu m_{\pi'\nu'}\, s_{\nu'}(Z)\cr
		             &= \sum_{k\ge 0}\, (\,s_{(1^k)}[s_{\pi}](Z)\,)' \sum_{k\ge 0}\, (-1)^{k|\pi|} ((-1)^k s_{(1^k)}[s_{\pi}(Z)])'\cr
								 &= \sum_\nu (-1)^{|\nu|}\,\ell_{\pi\nu}\, s_{\nu'}(Z) \nonumber 
\end{align}
so that $m_{\pi'\nu'}=(-1)^{|\nu|}\,\ell_{\pi\nu}$.

Applying these results to (\ref{eqn:s*piLM}) with $|\pi|$ even gives 
\begin{align}
    s^{*(\pi)}_\lambda(X) &=  [s_\lambda(Z)] \ L(XZ)\,L_{\pi'}(Z) = [s_\lambda(Z)]\ \prod_\mu\, (-1)^{|\mu|} s_\mu(X)\, s_{\mu'}(Z)\  \prod_\nu\, \ell_{\pi'\nu'}\, s_{\nu'}(Z) \cr
		                     &=  [s_\lambda(Z)] \prod_{\mu,\nu}\, (-1)^{|\mu|}\, \ell_{\pi'\nu'}\,s_\mu(X)\, \prod_\kappa\, c_{\mu'\nu'}^{\kappa}\, s_{\kappa}(Z) 
												               = \prod_{\mu,\nu}\, (-1)^{|\lambda|}\, \ell_{\pi'\nu'}\, c_{\mu\nu}^{\lambda'}\, s_\mu(X) \cr
												 &=  \prod_{\nu}\,(-1)^{|\lambda|} \ell_{\pi\nu}\,s_{\lambda'/\nu}(X) = (-1)^{|\lambda|}\, L^\perp_{\pi}(X) s_{\lambda'}(X)\,, \nonumber
\end{align}
where we have used the fact that $|\nu|$ is even, so that $(-1)^{|\mu|}=(-1)^{|\lambda|}$, and the fact that $c_{\mu'\nu'}^{\kappa}=c_{\mu\nu}^{\kappa'}$.
In the case $|\pi|$ odd we have
\begin{align}
    s^{*(\pi)}_\lambda(X) &=  [s_\lambda(Z)] \ L(XZ)\,M_{\pi'}(Z) = [s_\lambda(Z)]\ \prod_\mu\, (-1)^{|\mu|} s_\mu(X)\, s_{\mu'}(Z)\  \prod_\nu\, m_{\pi'\nu'}\, s_{\nu'}(Z) \cr
		                     &=  [s_\lambda(Z)] \prod_{\mu,\nu}\, (-1)^{|\mu|}\, m_{\pi'\nu'}\, s_\mu(X)\, \prod_\kappa\, c_{\mu'\nu'}^{\kappa}\, s_{\kappa}(Z) 
												                = \prod_{\mu,\nu}\, (-1)^{|\lambda|-|\nu|}\,m_{\pi'\nu'}\, c_{\mu\nu}^{\lambda'}\, s_\mu(X) \cr
												 &=  \prod_{\nu}\,(-1)^{|\lambda|} \ell_{\pi\nu}\,s_{\lambda'/\nu}(X) = (-1)^{|\lambda|}\, L^\perp_{\pi}(X)\, s_{\lambda'}(X)\,. \nonumber
\end{align}
Somewhat remarkably, these two formulae for $|\pi|$ even and odd coincide, and {yield (\ref{eqn:spi-Lperps}). The identification with
$\smash{s^{(\pi)}_{\lambda'}}(X)$ is evident in view of (\ref{eqn:SpiLambdaDef}). The forms
(\ref{eqn:s-Mperpspi}) follow trivially by left-multiplication by $M^\perp_{\pi}$, the inverse of $L^\perp_{\pi}$\,. This
completes the proof of our theorem.}
\qed

\section{Conclusions and related work.}
\label{sec:Conclusions}
The present paper gives further development of the algebraic-combinatorial context for the description of the general classes of symmetric functions of $\pi$-type, introduced in our previous papers \cite{fauser:jarvis:king:wybourne:2005a,fauser:jarvis:king:2010d,fauser:jarvis:king:2010b}. Specifically, further to the previously-derived $\pi$-type `plethystic' vertex operators given as the algebraic tools for deriving the $\pi$-type symmetric functions as modal products, we have identified in this work the dual counterparts of these objects. 
Quite surprisingly, our results generalize the well-known correspondence between standard vertex operator modes and their duals, and the general `free fermion' relations of the complex infinite Clifford algebra, to the case of the $\pi$-vertex operators and their duals. This is a significant extension, {endowing} the $\pi$-type symmetric functions with algebraic and combinatorial underpinnings which parallel those known for the Schur functions themselves. 

On the combinatorial side therefore, our results may inform constructions relating to determinantal forms for the $\pi$-type symmetric functions. Indeed, such combinatorial structures have been scrutinized by many
researchers over time, and it is beyond the scope of the present work to review
all of them. We mention\footnote{We thank a referee
for drawing our attention to this work.} 
one approach to the relationship between vertex operator realizations of  generalised symmetric functions and 
Jacobi-Trudi identities recently
discussed by Jing and Rozhkowskaya~\cite{Jing:Rozhkowskaya:2016}. 

Classically, Schur functions seen as
characters of the general linear group, can be constructed via
Jacobi-Trudi determinantal formulae, based on the parts of the
partition $\lambda$,
\begin{align*}
s_\lambda =&\, \det\left[
\begin{array}{c} h_{\lambda_i-i+j}\end{array}
\right]_{1\leq i,j\leq\ell(\lambda)}\,.
\end{align*}
Here the Schur function (character)
is expressed in terms of more basic objects, complete symmetric functions
$h_n$. In such cases,
different bases of symmetric functions, and such series as needed for some
restricted groups, can be produced by recurrence relations using an automorphism $\phi$ of the symmetric
function ring $\Lambda(X)$ of the form $u_{n+1} = \phi(u_{n})$, or  
multi-term or differential style extensions. The identification  $\smash{h^{(0)}_n} = h_n$, leads by iteration to 
$\smash{h^{(p)}_n} = \phi^{(p)}\big(\smash{h^{(0)}_{n}}\big) = h_{n+p}$ 
for a certain $\phi$. This effects a separation of row and column indices,
and recovers the Jacobi-Trudi formula via
\begin{align*}
s_\lambda &= \det\left[
\begin{array}{c} h^{(j-1)}_{\lambda_i-i+1}\end{array}
\right]_{1\leq i,j\leq\ell(\lambda)} 
\equiv
\det\left[
\begin{array}{c} h^{(0)}_{\lambda_i-i+j}\end{array}
\right]_{1\leq i,j\leq\ell(\lambda)}\,.
\end{align*}
We refer to~\cite{Jing:Rozhkowskaya:2016} for details of how characters of classical groups emerge as Jacob-Trudi
determinants for other definitions of $\smash{h^{(p)}_n}$ polynomials, and how
factorizations can be achieved for various bases. It should be noted that
not all vertex operators based on such polynomial sequences do allow Jacobi-Trudi determinants.
As one example that does so, for the orthogonal or symplectic algebras (\cite{Jing:Rozhkowskaya:2016}, Theorem 5.1) one has
\begin{align*}
\chi_{\mathfrak{g}}(\lambda) &= 
\det\left[
\begin{array}{c} h^{(j-1)}_{\lambda_i-i+1}\end{array}
\right]\,, \qquad
h^{(r)}_a = \left\{
\begin{array}{ll}
J_{a+r} + J_{a-r} & r > 0 \\
J_{a+r}           & r\leq 0
\end{array}\,,
\right.
\end{align*}
a result of Weyl~\cite{weyl:1930a}, see the text by Fulton and Harris~\cite{Fulton:Harris} (see also~\cite{Jing:Rozhkowskaya:2016}, Remark 5.2). The extension of these characters to the corresponding orthogonal or symplectic group coincides precisely with the $\pi$-Schur functions $\smash{s_\lambda^{(\pi)}}$ in the cases $\pi = (2)$ and $(1^2)$, respectively. It remains to be seen if the appropriate $\smash{h^{(p)}_n}$ can be identified
so as to yield a Jacobi-Trudi realisation of $\smash{s_\lambda^{(\pi)}}$ in the general case.

Unifying links to earlier work in this vein include for example Sergeev \cite{Sergeev}, and
Winkel \cite{Winkel1,Winkel2}. The work of Thomas \cite{Thomas}
used different methods to generate bases from
more elaborate algebraic and combinatorial tools, such as domino tableaux and
Baxter operators, and addressed the stability property (pro-finiteness)
of such polynomial sequences. An abstract, Hopf-algebraic
deformation theory was subsequently developed by Rota and Stein
\cite{Rota-Stein1,Rota-Stein2}. 

A detailed reconciliation of the categorical, Hopf-algebra inspired, but in principle basis-free, methods underlying our present derivations (as in~\cite{fauser:jarvis:2003a,fauser:jarvis:king:wybourne:2005a,fauser:jarvis:king:2010b}), with the adapted bases of Jing and Rozhkowskaya~\cite{Jing:Rozhkowskaya:2016}, better matched to the physical point of view using creation and annihilation of composite or effective particles, is beyond the scope of this paper, and a subject for further investigation\footnote{After this paper was completed, we became aware of the work~\cite{Bytsenko:Chaichian:2016} which
links formal characters of the classical groups and vertex operator expressions to certain homology 
invariants in hyperbolic geometry. The extension of these considerations to
encompass also the role of the $\pi$-type symmetric functions is equally a topic for future study.}.

As a final direction emerging from the present work, we note the well-known connection between the formal fermion-boson mapping \cite{jimbo:miwa:1983a} and various infinite hierarchies of solitonic equations such as the KP hierarchy. As has already been explored by Baker~\cite{baker1996vertex} for the orthogonal and symplectic cases, it is possible to appropriate the classical method of coordinatizing the vacuum orbit of suitably defined elements of $GL(\infty)$ -- realized as vertex operators -- to recover alternative hierarchies in Hirota bilinear form with, amongst other properties, symmetric functions of orthogonal and symplectic type as tau functions  (see also
~\cite{Baker:Jarvis:Yung:1993,Jarvis:Yung1993}). The extension to the case of generalized symmetric functions of $\pi$ type will be developed in a future paper.

\appendix

\section{{Reordering relations}}
\renewcommand{\theequation}{A--\arabic{equation}}
\setcounter{equation}{0}
\label{sec:Appendix A}

\begin{Lemma}\label{lem:reordering}
{For all partitions $\pi$ and any $z$ and $w$,} we have the following reordering identities:
\begin{align}
  M^\perp_\pi(z) M(w) &=  
	                        M(w)\  \prod_{k\ge 0}\, M^\perp_{\pi/(k)}(zw^k) \,;  \label{eqn:MpiperpM} \\
  L^\perp_\pi(z) M(w) &=   
                          M(w)\  \prod_{k\ge 0} \, L^\perp_{\pi/(k)}(zw^k) \,; \label{eqn:LpiperpM} \\
  M^\perp_\pi(z) L(w) &=  
	                        L(w)\  \prod_{k\ge 0}\, M^\perp_{\pi/(1^{2k})}(zw^{2k})\, L^\perp_{\pi/(1^{2k+1})}(zw^{2k+1}) \,; \label{eqn:MpiperpL} \\
  L^\perp_\pi(z) L(w) &=  
	                        L(w)\  \prod_{k\ge 0}\, L^\perp_{\pi/(1^{2k})}(zw^{2k})\, M^\perp_{\pi/(1^{2k+1})}(zw^{2k+1}) \,. \label{eqn:LpiperpL}
\end{align}
\end{Lemma}

\noindent{\bf Proof}:

{In order to establish the required reordering relations for skew Schur function series it is helpful to
note some properties of plethysms and skew Schur functions. First,}
let $\rho$ and $\xi$ be partitions of $r$ and $k$ respectively. Then for any alpahabet $X=(x_1,x_2,\ldots)$
\begin{align}
    s_\rho[s_\xi(X)] \quad\hbox{contains}\quad s_{(kr)}(X) &\Leftrightarrow \hbox{$\rho=(r)$ and $\xi=(k)$}; \label{eqn:pleth-kr}\\ \cr
		s_\rho[s_\xi(X)] \quad\hbox{contains}\quad s_{(1^{kr})}(X) &\Leftrightarrow \begin{cases} \hbox{$\rho=(r)$ and $\xi=(1^k)$ with $k$ even, \, or},\cr
		                                                                                 \hbox{$\rho=(1^r)$ and $\xi=(1^k)$ with $k$ odd}.\cr
																																										\end{cases} \label{eqn:pleth-1kr}
\end{align}
Each of $s_{(kr)}(X)$ and $s_{(1^{kr})}(X)$, if it occurs, has multiplicity 1. It follows that
\begin{align}
    s_{(m)}(X) / s_\rho[s_\xi(X)] &= \begin{cases} 
		                                  s_{(m-kr)}(X)&\hbox{if $\rho=(r)$ and $\xi=(k)$}\,; \cr 
		                                              0&\hbox{otherwise}\,,\cr
																		  \end{cases} \label{eqn:m-skew-pleth} \\
		s_{(1^m)}(X) / s_\rho[s_\xi(X)] &= \begin{cases}
		                           s_{(1^{m-kr})}(X)&\hbox{if} \begin{cases} \hbox{$\rho=(r)$ and $\xi=(1^k)$ with $k$ even, \, or}\cr
		                                                                     \hbox{$\rho=(1^r)$ and $\xi=(1^k)$ with $k$ odd}\,;.\cr
																													\end{cases} \cr
															         0&\hbox{otherwise}\,,\cr                                                     
																	      \end{cases}  \label{eqn:1m-skew-pleth} 
\end{align}
with $s_{(m-kr)}(X)=s_{(1^{m-kr})}(X)=0$ if $kr>m>$.

{The coproduct skew action of each of our $\pi$-Schur function series has been described elsewhere~\cite{fauser:jarvis:king:wybourne:2005a}.} 
The skew quotient of $M(w,X) \times G(Y)$ by $M_\pi(z;X,Y)$ for any symmetric function $G(Y)$ may then be evaluated as follows, where the dependence on $X$ and $Y$ has been suppressed:
\begin{align}
  &(\,M(w) \times G\,) / M_\pi(z) =  \prod_{(0)\subseteq\xi\subseteq\pi} \sum_{\rho} z^{|\rho|}\ (\, M(w) / s_\rho[s_\xi] \,) \times (\,G / s_\rho[s_{\pi/\xi}]\,)\cr
	                &=  \prod_{(0)\subseteq\xi\subseteq\pi} \sum_{\rho} z^{|\rho|} \sum_{m\geq0} w^m\,s_{(m-kr)}\,\delta_{\rho,(r)}\,\delta_{\xi,(k)} \times ( G / s_\rho[s_{\pi/\xi}] )\cr
									&=  \prod_{(0)\subseteq(k)\subseteq\pi} \sum_{r\geq0} z^r \sum_{m\geq0} w^m\,s_{(m-kr)} \times ( G / s_{(r)}[s_{\pi/(k)}] )\cr
									&=  \prod_{(0)\subseteq(k)\subseteq\pi} \sum_{r\geq0} z^r \sum_{n\geq0} w^{n+kr}\,s_{(n)} \times ( G / s_{(r)}[s_{\pi/(k)}] )\cr
									&=  \sum_{n\geq0} w^n\, s_{(n)} \times G \,\big/\, \!\!\!\! \prod_{(0)\subseteq(k)\subseteq\pi} \sum_{r\geq0} (zw^k)^r\, s_{(r)}[s_{\pi/(k)}]	\cr
						      &=  \ M(w) \times G \,\big/\,  \prod_{k\ge 0}\, M_{\pi/(k)}(zw^k)\,, \label{eqn:MGskewMpi}
\end{align}
{where (\ref{eqn:MLequal0}) and (\ref{eqn:MLequal1}) justify the last step}.
Similarly
\begin{align}
  &(\,M(w) \times G\,) / L_\pi(z) =  \prod_{(0)\subseteq\xi\subseteq\pi} \sum_{\rho}(-z)^{|\rho|}\ (\, M(w) / s_\rho[s_\xi] \,) \times (\,G / s_{\rho'}[s_{\pi/\xi}]\,)\cr
	                &=  \prod_{(0)\subseteq\xi\subseteq\pi} \sum_{\rho} (-z)^{|\rho|} \sum_{m\geq0} w^m\,s_{(m-kr)}\,\delta_{\rho,(r)}\,\delta_{\xi,(k)} \times ( G / s_{\rho'}[s_{\pi/\xi}] )\cr
									&=  \prod_{(0)\subseteq(k)\subseteq\pi} \sum_{r\geq0} (-z)^r \sum_{m\geq0} w^m\,s_{(m-kr)} \times ( G / s_{(1^r)}[s_{\pi/(k)}] )\cr
									&=  \prod_{(0)\subseteq(k)\subseteq\pi} \sum_{r\geq0} (-z)^r \sum_{n\geq0} w^{n+kr}\,s_{(n)} \times ( G / s_{(1^r)}[s_{\pi/(k)}] )\cr
									&=  \sum_{n\geq0} w^n\, s_{(n)} \times G \,\big/\, \!\!\!\!  \prod_{(0)\subseteq(k)\subseteq\pi} \sum_{r\geq0} (-zw^k)^r\, s_{(1^r)}[s_{\pi/(k)}] \cr
						      &=\  M(w) \times G \,\big/\,  \prod_{k\ge 0}\, L_{\pi/(k)}(zw^k) \,. \label{eqn:MGskewLpi}
\end{align}
While
\begin{align}
  &(\,L(w) \times G\,) / M_\pi(z) =  \prod_{(0)\subseteq\xi\subseteq\pi} \sum_{\rho} z^{|\rho|}\ (\, L(w) / s_\rho[s_\xi] \,) \times (\,G / s_\rho[s_{\pi/\xi}]\,)\cr
	                &=  \prod_{(0)\subseteq\xi\subseteq\pi} \sum_{\rho} z^{|\rho|} \sum_{m\geq0} (-w)^m\, s_{(1^{m-kr})}\,\delta_{\xi,(1^k)} 
									     \left(\,\delta_{\rho,(r)}\,\chi(k\,\mbox{even})\,+\,\delta_{\rho,(1^r)}\,\chi(k\,\mbox{odd})\, \right)
									            \times ( G / s_\rho[s_{\pi/\xi}] )\cr
									&=  \prod_{(0)\subseteq(1^{k})\subseteq\pi} \sum_{r\geq0} z^r \sum_{n\geq0} (-w)^{n+kr}\,s_{(1^{n})} 
											\times G \,\big/\,  \left(\,s_{(r)}[s_{\pi/(1^{k})}] \,\chi(k\,\mbox{even}) \,+\, s_{(1^r)}[s_{\pi/(1^{k})}] \,\chi(k\,\mbox{odd})\, \right) \cr
									&=  \sum_{n\geq0}(-w)^n\, s_{(n)} \times G \,\big/\,\!\!\!\! \prod_{(0)\subseteq(1^{2k})\subseteq\pi} \sum_{r\geq0} (zw^{2k})^r\, s_{(r)}[s_{\pi/(1^{2k})}]
								               \!\!\!\! \prod_{(0)\subseteq(1^{2k+1})\subseteq\pi} \sum_{r\geq0} (-zw^{2k+1})^r\, s_{(1^r)}[s_{\pi/(1^{2k+1})}] )\cr
									&=  L(w) \times G \,\big/\,  \prod_{k\ge 0}\, M_{\pi/(1^{2k})}(zw^{2k})\
									                    \prod_{k\ge 0}\, L_{\pi/(1^{2k+1})}(zw^{2k+1}) \,. \label{eqn:LGskewMpi}
\end{align}
and 
\begin{align}
  &(\,L(w) \times G\,) / L_\pi(z) =  \prod_{(0)\subseteq\xi\subseteq\pi} \sum_{\rho}(-z)^{|\rho|}\ (\, L(w) / s_\rho[s_\xi] \,) \times (\,G / s_{\rho'}[s_{\pi/\xi}]\,)\cr
	                &=  \prod_{(0)\subseteq\xi\subseteq\pi} \sum_{\rho} (-z)^{|\rho|} \sum_{m\geq0} (-w)^m\, s_{(1^{m-kr})}\,\delta_{\xi,(1^k)} 
									     \left( \,\delta_{\rho,(r)}\,\chi(k\,\mbox{even})\,+\,\delta_{\rho,(1^r)}\,\chi(k\,\mbox{odd})\, \right)
									            \times ( G / s_{\rho'}[s_{\pi/\xi}] )\cr
									&=  \prod_{(0)\subseteq(1^{k})\subseteq\pi} \sum_{r\geq0} (-z)^r \sum_{n\geq0} (-w)^{n+kr}\,s_{(1^{n})} 
									      \times G \,\big/\,  \left(\,s_{(1^r)}[s_{\pi/(1^{k})}] \,\chi(k\,\mbox{even}) \,+\, s_{(r)}[s_{\pi/(1^{k})}] \,\chi(k\,\mbox{odd})\, \right) \cr
									&=  \sum_{n\geq0}(-w)^n\, s_{(n)} \times G \,\big/\, \!\!\!\!  \prod_{(0)\subseteq(1^{2k})\subseteq\pi} \sum_{r\geq0} (-zw^{2k})^r\, s_{(1^r)}[s_{\pi/(1^{2k})}] )
									         \!\!\!\! \prod_{(0)\subseteq(1^{2k+1})\subseteq\pi} \sum_{r\geq0} (-zw^{2k+1})^r\, s_{(r)}[s_{\pi/(1^{2k+1})}] )\cr
									&=  L(w) \times G \,\big/\, \prod_{k\ge 0}\, L_{\pi/(1^{2k})}(zw^{2k})
									                    \prod_{k\ge 0}\, M_{\pi/(1^{2k+1})}(zw^{2k+1}) \,. \label{eqn:LGskewLpi}
\end{align}
{where $\chi$ is the truth function so that $\chi(k\,\mbox{even})=1$ if $k$ is even and $=0$ if $k$ is odd, while
$\chi(k\,\mbox{odd})=1$ if $k$ is odd and $=0$ if $k$ is even.}

{To complete the proof} we simply re-write the above skew coproduct relations as operator statements without explicit reference to the arbitrary $G$ series.
\qed

\section{{Reordering zero mode modifiers}}
\renewcommand{\theequation}{B--\arabic{equation}}
\setcounter{equation}{0}
\label{sec:Appendix B}
\noindent
As mentioned in the text, to the respective vertex operators $V_\pi(z)$, $V^*_\pi(z)$ are appended the contributions
from zero modes, which are {associated with the} canonically conjugate operators $\alpha_0$ and $q$. For clarity we 
give here suitable ordered forms
showing which additional $z,w$-dependent factors are needed to go from the anticommutation relations for $V_\pi $, $V^*_\pi $ 
to the free fermion anticommutation relations satisfied by the full vertex operators  $X^\pi$, $X^{*\pi}$\,. 

\begin{Lemma}\label{lem:zero_mode_factors}
\begin{align}
V_\pi(z)V_\pi(w):\qquad  &
e^{iq}z^{\alpha_0}e^{iq}w^{\alpha_0} = (1/zw^2)\, (zw)^{\alpha_0}\, e^{2iq}\,; \label{eqn:XXzw} \\
V^*_\pi(z)V^*_\pi(w):\qquad &
z^{-\alpha_0}e^{-iq}w^{-\alpha_0}e^{-iq} = (1/w)\, (zw)^{-\alpha_0}\, e^{-2iq}\,;  \label{eqn:X*X*zw} \\
V_\pi(z)V^*_\pi(w):\qquad &
e^{iq}z^{\alpha_0}w^{-\alpha_0}e^{-iq} = (w/z)\,(z/w)^{\alpha_0}\,;  \label{eqn:XX*zw} \\
V^*_\pi(w)V(z)_\pi:\qquad &
w^{-\alpha_0}e^{-iq}e^{iq}z^{\alpha_0} = (z/w)^{\alpha_0}\,. \label{eqn:X*Xwz}
\end{align}
\end{Lemma}

\noindent{\bf Proof}:
For all $z$ it should be noted that thanks to (\ref{eqn:CauchyExpFormula}) we have
\begin{align}
 e^{iq}\, z^{\alpha_0} &= e^{iq}\, e^{(ln z)\alpha_0} = e^{(\ln z)\alpha_0} e^{-(\ln z)\alpha_0} e^{iq}\, e^{(\ln z)\alpha_0} \cr
                       &= z^{\alpha_0}\,  e^{iq + [iq,(\ln z)\alpha_0] +\cdots} = z^{\alpha_0}\,  e^{iq + i(\ln z)[q,\alpha_0] +\cdots}\cr 
											 &= z^{\alpha_0}\,  e^{iq - (\ln z)} =  (1/z)\, z^{\alpha_0}\,  e^{iq}\,, 
\end{align}
since $[q,\alpha_0]=i$. Similarly, 
\begin{equation}
 e^{-iq}\, z^{-\alpha_0} =  (1/z)\, z^{-\alpha_0}\,  e^{-iq}\,
\end{equation}
These reordering relations are sufficient to establish very easily the validity of (\ref{eqn:XXzw})-(\ref{eqn:X*Xwz}).
\qed
\medskip

\section{{Products of vertex operators and their duals}}
\renewcommand{\theequation}{C--\arabic{equation}}
\setcounter{equation}{0}
\label{sec:Appendix C}

In order to prove Theorem~\ref{thm:spi-s*pi} it is necessary to make use of the following
vertex operator reordering lemma:
\begin{Lemma}\label{lem:VVV*V*}
Let $Z=(z_1,z_2,\ldots,z_m)$, then for all partitions $\pi$ and suppressing the dependence on $X=(x_1,x_2,\ldots)$ we have
\begin{align}
    V_\pi(z_1) V_\pi(z_2) \cdots V_\pi(z_m) &= \prod_{1\le i<j\le m} (1-\ov{z_i}z_j)\ \prod_{\ell=1}^m\, M(z_\ell)\, L^\perp(\ov{z_\ell}) \cr
		                 &~~~~ \cdot   \prod_{\begin{array}{c} \sc i_1,i_2,\ldots,i_m\geq 0\cr \sc (i_1,i_2,\ldots,i_m)\neq(0,0,\ldots,0) \cr \end{array}} 
													 L^\perp_{\pi/((i_1)(i_2)\cdots(i_m))} (z_1^{i_1}z_2^{i_2}\cdots z_m^{i_m})\,; \label{eqn:VVV} \\
		 V^*_\pi(z_1) V^*_\pi(z_2) \cdots V^*_\pi(z_m) &= \prod_{1\le i<j\le m} (1-\ov{z_i}z_j)\ \prod_{\ell=1}^m\, L(z_\ell)\, M^\perp(\ov{z_\ell}) \cr
		                 &~~~~ \cdot   \prod_{\begin{array}{c} \sc i_1,i_2,\ldots,i_m\geq 0\cr \sc i_1+i_2+\cdots+i_m \mbox{odd} \cr \end{array}} 
													 M^\perp_{\pi/((1^{i_1})(1^{i_2})\cdots(1^{i_m}))} (z_1^{i_1}z_2^{i_2}\cdots z_m^{i_m}) \cr		
		              &~~~~ \cdot \hskip-1ex \prod_{\begin{array}{c} \sc i_1,i_2,\ldots,i_m\geq 0\cr \sc i_1+i_2+\cdots+i_m \mbox{even} \cr \sc (i_1,i_2,\ldots,i_m)\neq(0,0,\ldots,0) \cr \end{array}} 
													 L^\perp_{\pi/((1^{i_1})(1^{i_2})\cdots(1^{i_m}))} (z_1^{i_1}z_2^{i_2}\cdots z_m^{i_m})\,. \label{eqn:V*V*V*} 							
\end{align}
\end{Lemma}
 
\noindent{\bf Proof}:
We proceed by induction with respect to $m$. 
The case $m=1$ is covered by the definitions (\ref{eqn:Vpi}) and (\ref{eqn:V*pi}),
while the case $m=2$ was proved in the first two parts of the proof of Theorem~\ref{thm:main} through
the derivation of the results (\ref{eqn:VpiVpi})-(\ref{eqn:P}) and (\ref{eqn:V*piV*pi})-(\ref{eqn:Q}). 

By hypothesis, let the results be true for $Z'=(z_1,z_2,\ldots,z_{m-1})$ and consider the case $Z=(z_1,z_2,\ldots,z_m)$.
In the case of (\ref{eqn:VVV}) this gives
\begin{align}
     & V_\pi(z_1) V_\pi(z_2) \cdots V_\pi(z_m) \cr
		&= \prod_{1\le i<j\le m-1} (1-\ov{z_i}z_j)\ \prod_{\ell=1}^{m-1}\, M(z_\ell)\, L^\perp(\ov{z_\ell}) 
		  \!\!\!\!\!\! \prod_{\begin{array}{c} \sc i_1,i_2,\ldots,i_{m-1}\geq 0\cr \sc (i_1,i_2,\ldots,i_{m-1})\neq(0,0,\ldots,0) \cr \end{array}} 
													\!\!\!\!\!\! L^\perp_{\pi/((i_1)(i_2)\cdots(i_{m-1}))}(z_1^{i_1}z_2^{i_2}\cdots z_{m-1}^{i_{m-1}}) \cr
		&~~~~\cdot~~~~ M(z_m)\, L^\perp(\ov{z_m})\ \prod_{i_m>0} L^\perp_{\pi/(i_m)}(z_m^{i_m}) \\,. \nonumber
\end{align}
However
\[
    L^\perp(\ov{z_\ell})\, M(z_m) = M(z_m) \, L^\perp(\ov{z_\ell})\, L^\perp_{(0)}(\ov{z_\ell}z_m) 
		   = (1-\ov{z_\ell}z_m)\, M(z_m)\, L^\perp(\ov{z_\ell}) \,, \nonumber							
\] 
and 
\[
   L^\perp_{\pi/((i_1)(i_2)\cdots(i_{m-1}))} (z_1^{i_1}z_2^{i_2}\cdots z_{m-1}^{i_{m-1}}) \, M(z_m) 
	  = M(z_m) \prod_{i_m\ge 0}\,  L^\perp_{\pi/((i_1)(i_2)\cdots(i_m))} (z_1^{i_1}z_2^{i_2}\cdots z_{m}^{i_{m}}) 
\]
It follows that
\begin{align}
      &V_\pi(z_1) V_\pi(z_2) \cdots V_\pi(z_m) = 
			                                     \prod_{0\le i<j\le m} (1-\ov{z_i}z_j)\ \prod_{\ell=0}^m\, M(z_\ell)\, L^\perp(\ov{z_\ell}) \cr
		  &~~~~\cdot~~~~ \prod_{\begin{array}{c} \sc i_1,i_2,\ldots,i_m\geq 0\cr \sc (i_1,i_2,\ldots,i_{m-1})\neq(0,0,\ldots,0) \cr \end{array}} 
													 L^\perp_{\pi/((i_1)(i_2)\cdots(i_m))} (z_1^{i_1}z_2^{i_2}\cdots z_m^{i_m}) 
											\		\prod_{i_m>0} L^\perp_{\pi/(i_m)}(z_m^{i_m}) 									\,. \nonumber
\end{align}
This is precisely the required (\ref{eqn:VVV}), thereby completing the induction argument.

The argument in the case of (\ref{eqn:V*V*V*}) is more intricate, but again by the induction hypothesis one assumes the
validity of the result involving $Z'=(z_1,z_2,\ldots,z_{m-1})$ and then multiplies by an additional factor $V^*_\pi(z_m)$. This gives
\begin{align}
      &V^*_\pi(z_1) V^*_\pi(z_2) \cdots V^*_\pi(z_m) = 
						           \prod_{1\le i<j\le {m-1}} (1-\ov{z_i}z_j)\ \prod_{\ell=1}^{m-1}\, L(z_\ell)\, M^\perp(\ov{z_\ell}) \cr
		  &~~~\cdot~~~\prod_{\begin{array}{c} \sc i_1,i_2,\ldots,i_{m-1}\geq 0\cr \sc i_1+i_2+\cdots+i_{m-1} odd \cr \end{array}} 
									 M^\perp_{\pi/((1^{i_1})(1^{i_2})\cdots(1^{i_{m-1}}))} (z_1^{i_1}z_2^{i_2}\cdots z_{m-1}^{i_{m-1}}) \cr
		  &~~~\cdot~~~\prod_{\begin{array}{c} \sc i_1,i_2,\ldots,i_{m-1}\geq 0\cr \sc i_1+i_2+\cdots+i_{m-1} even \cr \sc(i_1,i_2,\ldots,i_{m-1})\neq(0,0,\ldots,0) \cr \end{array}} 
									 L^\perp_{\pi/((1^{i_1})(1^{i_2})\cdots(1^{i_{m-1}}))} (z_1^{i_1}z_2^{i_2}\cdots z_{m-1}^{i_{m-1}})  \cr
			&~~~\cdot~~~ L(z_m)\, M^\perp(\ov{z_m})\, \prod_{i_m\ge 0;i_m odd}\, M^\perp_{\pi/(1^{i_m})}(z_m^{i_m})\ \prod_{i_m\ge 0;i_m even;i_m\neq 0}\, L^\perp_{\pi/(1^{i_m})}(z_m^{i_m}) 						\,. \nonumber
\end{align}
This time we have 
\[
    M^\perp(\ov{z_\ell}) \, L(z_m) = L(z_m) \, M^\perp(\ov{z_\ell})\, L^\perp_{(0)}(\ov{z_\ell}z_m) 
		   = (1-\ov{z_\ell}z_m)\, L(z_m) M^\perp(\ov{z_\ell}) \,,						
\] 
while
\begin{align}
      &M^\perp_{\pi/((1^{i_1})(1^{i_2})\cdots(1^{i_{m-1}}))} (z_1^{i_1}z_2^{i_2}\cdots z_{m-1}^{i_{m-1}}) \ L(z_m)  \cr
			&= L(z_m) \, \prod_{i_m\ge 0; i_m even}\, M^\perp_{\pi/((1^{i_1})(1^{i_2})\cdots(1^{i_m}))} (z_1^{i_1}z_2^{i_2}\cdots z_{m}^{i_{m}}) \cr
			&~~~~~~~~~~~~~~~~~~~ \prod_{i_m\ge 0; i_m odd}\, L^\perp_{\pi/((1^{i_1})(1^{i_2})\cdots(1^{i_m}))} (z_1^{i_1}z_2^{i_2}\cdots z_{m}^{i_{m}}) \,,
\end{align}
and
\begin{align}
      &L^\perp_{\pi/((1^{i_1})(1^{i_2})\cdots(1^{i_{m-1}}))} (z_1^{i_1}z_2^{i_2}\cdots z_{m-1}^{i_{m-1}}) \ L(z_m)  \cr
			&= L(z_m) \, \prod_{i_m\ge 0; i_m even}\, L^\perp_{\pi/((1^{i_1})(1^{i_2})\cdots(1^{i_m}))} (z_1^{i_1}z_2^{i_2}\cdots z_{m}^{i_{m}}) \cr
			&~~~~~~~~~~~~~~~~~~~ \prod_{i_m\ge 0; i_m odd}\, M^\perp_{\pi/((1^{i_1})(1^{i_2})\cdots(1^{i_m}))} (z_1^{i_1}z_2^{i_2}\cdots z_{m}^{i_{m}}) \,.
\end{align}
Hence
\begin{align}
      &V^*_\pi(z_1) V^*_\pi(z_2) \cdots V^*_\pi(z_m) = 
						           \prod_{1\le i<j\le {m}} (1-\ov{z_i}z_j)\ \prod_{\ell=1}^{m}\, L(z_\ell)\, M^\perp(\ov{z_\ell}) \cr
		  &~~~\cdot~~~\prod_{\begin{array}{c} \sc i_1,i_2,\ldots,i_{m}\geq 0\cr \sc i_1+i_2+\cdots+i_{m-1} odd, i_m even \cr \end{array}} 
									 M^\perp_{\pi/((1^{i_1})(1^{i_2})\cdots(1^{i_{m-1}}))} (z_1^{i_1}z_2^{i_2}\cdots z_{m}^{i_{m}}) \cr
			&~~~\cdot~~~\prod_{\begin{array}{c} \sc i_1,i_2,\ldots,i_{m}\geq 0\cr \sc i_1+i_2+\cdots+i_{m-1} odd, i_m odd \cr \end{array}}
									 L^\perp_{\pi/((1^{i_1})(1^{i_2})\cdots(1^{i_{m}}))} (z_1^{i_1}z_2^{i_2}\cdots z_{m}^{i_{m}}) \cr 
		  &~~~\cdot~~~\prod_{\begin{array}{c} \sc i_1,i_2,\ldots,i_{m}\geq 0\cr \sc i_1+i_2+\cdots+i_{m-1} even, i_m even \cr 
			                                    \sc(i_1,i_2,\ldots,i_{m-1})\neq(0,0,\ldots,0) \cr \end{array}} 
									 L^\perp_{\pi/((1^{i_1})(1^{i_2})\cdots(1^{i_{m}}))} (z_1^{i_1}z_2^{i_2}\cdots z_{m}^{i_{m}}) \cr
			&~~~\cdot~~~\prod_{\begin{array}{c} \sc i_1,i_2,\ldots,i_{m}\geq 0\cr \sc i_1+i_2+\cdots+i_{m-1} even, i_m odd \cr 
									                        \sc(i_1,i_2,\ldots,i_{m-1})\neq(0,0,\ldots,0) \cr \end{array}} 
									 M^\perp_{\pi/((1^{i_1})(1^{i_2})\cdots(1^{i_{m}}))} (z_1^{i_1}z_2^{i_2}\cdots z_{m}^{i_{m}}) \cr
			&~~~\cdot~~~\prod_{i_m\ge 0;i_m odd}\, M^\perp_{\pi/(1^{i_m})}(z_m^{i_m}) \, \prod_{i_m\ge 0;i_m even;i_m\neq 0}\, L^\perp_{\pi/(1^{i_m})}(z_m^{i_m})
															\,. \nonumber
\end{align}
As before this is precisely the required (\ref{eqn:V*V*V*}), thereby completing the induction argument.
\qed

\bigskip

\noindent
\textbf{Acknowledgements}\\
The authors acknowledge the support of the Mathematisches Forschungsinstitut Oberwolfach for a `Research in Pairs' residence
in 2014, during which this work was developed. We thank the staff and management of MFO for their kind hospitality and
for the use of the MFO facilities during our stay.

\end{document}